\documentclass[structabstract]{aa}
\usepackage{txfonts}
\usepackage{graphicx}
\usepackage{natbib}
\bibpunct{(}{)}{;}{a}{}{,} 

\begin{document}

\title{Analyzing low signal-to-noise FUSE spectra:}
\subtitle{Confirmation of Lyman continuum escape from Haro 11}

\author{$^1$Elisabet Leitet
\and $^1$Nils Bergvall
\and $^1$Nikolai Piskunov 
\and $^2$B-G Andersson}

\institute{{Department of Physics and Astronomy, Uppsala University, Box 515, SE-751 20 Uppsala,
SWEDEN}\\
\email{elisabet.leitet@fysast.uu.se, nils.bergvall@fysast.uu.se, nikolai.piskunov@fysast.uu.se}
\and{Stratospheric Observatory for Infrared Astronomy, NASA Ames Research Center, Mail Stop 211-3, Moffett Field, CA 94035, USA}\\
\email{bgandersson@sofia.usra.edu}}

\date{Accepted: 26 May 2011 }

\abstract{Galaxies are believed to be the main providers of Lyman continuum (LyC) photons during the early phases of the cosmic reionization. Little is known however, when it comes to escape fractions and the mechanisms behind the leakage. To learn more, one may look at local objects, but so far only one low-z galaxy has shown any signs of emitting LyC radiation. With data from the Far Ultraviolet Spectroscopic Explorer (FUSE), we previously found an absolute escape fraction of ionizing photons ($f_{esc}$) of 4-10$\%$ for the blue compact galaxy \object{Haro 11}. However, using a revised version of the reduction pipeline on the same data set, Grimes and collaborators were unable to confirm this and derived an upper limit of $f_{esc}$$\lesssim$2$\%$ .}
{We attempt to determine whether Haro 11 is emitting ionizing radiation to a significant level or not. We also investigate the performance of the reduction pipeline for faint targets such as Haro 11, and introduce a new approach to the background subtraction.}
{The final version of the reduction pipeline, CalFUSE v3.2, was applied to the same Haro 11 data set as the two previous authors used. At these faint flux levels, both FUSE and CalFUSE are pushed to their limits, and a detailed analysis was undertaken to monitor the performance of the pipeline. We show that non-simultaneous background estimates are insuffient when working with data of low signal-to-noise ratio (S/N), and a new background model was developed based on a direct fit to the detector response.}
{We find that one has to be very careful when using CalFUSE v3.2 on low S/N data, and especially when dealing with sources where signal might originate from off-center regions. Applying the new background fit, a significant signal is detected in the LyC in both detector segments covering these wavelengths. Thus, the leakage is confirmed with a flux density of $f_{900}$=4.0 $\times$ 10$^{-15}$ erg s$^{-1}$ cm$^{-2}$ \AA$^{-1}$ (S/N=4.6),  measured on the airglow free regions in the LyC for the night-only data. This corresponds to an absolute escape fraction of ionizing photons from Haro 11 of $f_{esc}$=3.3$\pm0.7$ $\%$. We confirm these results by investigating the two-dimensional data, the count rates, and the residual flux in \ion{C}{ii} $\lambda$1036{\AA}. }
{}


\maketitle

\section{Introduction}
One of the main events in the history of the universe was the phase transition when the neutral gas again became ionized. Finding out when and how this cosmic reionization occured remains a challenge for observers in a broad range of fields. Observational constraints based on the seven-year WMAP data indicate that the reionization started at a redshift z$\approx$10.5 \citep{2010arXiv1001.4635L}. Over time, the intergalactic medium became ionized over larger and larger regions, but the redshift at which this process was completed is uncertain.  While some studies of the Gunn-Peterson effect, the evolution of the Ly$\alpha$ escape fraction, and the Ly$\alpha$ luminosity function indicate that reionization ended at z$\approx$6 (e.g. \citealt{2004ApJ...617L...5M}, \citealt{2006ARA&A..44..415F}, \citealt{2006PASJ...58..485T}, \citealt{2010arXiv1004.0963O}), other investigations give less support for this (e.g. \citealt{2010arXiv1006.3071T}, \citealt{2010MNRAS.tmp.1017M}, \citealt{2010arXiv1007.2961O}). A highly inhomogeneous intergalactic medium complicates the interpretation of the scanty observational data but if we assume that the reionization epoch took place over the redshift interval z=6-11, we can try to identify the sources generating the cosmological background of ionizing photons. Probably the most significant sources varied over time. At late times, z$<$3, it has been found that the bulk of the ionizing photons originates from quasars \citep{2008ApJ...688...85F} and AGNs \citep{2009ApJ...692.1476C}, while at higher redshifts quasars are too few in number to be the dominant source \citep{2008AJ....135.1057J,2009JCAP...03..022L}. The onset of the reionization has instead been connected to the appearance of the first luminous structures in the universe, the first stars and galaxies \citep{2001ApJ...549L.151H}. 

 It still remains unclear whether there were enough galaxies around at the epoch of reionization to account for the required number of ionizing photons, but lately several high-z surveys provide support for an early galaxy build-up.  Using the HST WFC3/IR camera, \citet{2009arXiv0912.4263B} derived the galaxy luminosity function for z $\approx$ 10 galaxies, and found an increasing likelihood of a galaxy-driven reionization. The evolution of the luminosity function together with models show that the onset of reionization must have been initiated by low mass galaxies (e.g. \citealt{2010ApJ...714L.202T,2010PASA...27..110S}), with properties similar to the observed low luminosity population at z$\sim$7 \citep{2010ApJ...708L..69B}. Cosmological simulations including stellar feedback (e.g. \citealt{2010ApJ...710.1239R,2010arXiv1002.3346Y}) give further support to the importance of LyC leakage from dwarf galaxies (halo masses M$_h\approx10^9$ M$_{\odot}$). The same models show a decreasing contribution from galaxies of increasingly higher mass (but see \citealt{2008ApJ...672..765G}).

Direct measurements of escaping ionizing photons from the galaxies that reionized the universe is unfortunately  impeded by the still highly opaque neutral intergalactic medium (IGM) at these redshifts. To measure escape fractions and investigate the physical properties that allow leakage, we need to study more nearby objects. This can and has been done with groundbased telescopes at z$\sim$3 (e.g.  \citealt{2001ApJ...546..665S, 2006ApJ...651..688S, 2009ApJ...692.1287I, 2010ApJ...725.1011V, 2011arXiv1102.0286N}), or locally by using space telescopes such as HUT \citep{1995ApJ...454L..19L}, FUSE \citep{2001A&A...375..805D,2006A&A...448..513B,2007ApJ...668..891G, 2009ApJS..181..272G}, GALEX \citep{2009ApJ...692.1476C}, and HST \citep{2003ApJ...598..878M, 2007ApJ...668...62S, 2010ApJ...723..241S, 2010ApJ...720..465B}.  Of the four only HST and GALEX are still in operation, and until now only HUT and FUSE have been providing wavelength coverage down to 900 \AA, such that local galaxies can be studied. With the successful installation of the new HST instrument COS, this wavelength region is yet again opened up. Despite all of these efforts, only a few investigations have resulted in LyC detections for single sources. There have been $\sim$50 single detections so far, all except one at z$\sim$3, although foreground contamination is expected to account for some of those. The one exception is the disputed case of Haro 11.

Over the years, FUSE has provided a wealth of information about the ultraviolet (UV) spectrum of stars and galaxies and made an outstanding contribution to UV astronomy. It was however not optimized for observing faint objects, and its sensitivity is at its lowest at the wavelengths that are ideal for measuring the Lyman continuum from local galaxies. Therefore, even though the FUSE archive still provides a unique legacy to aid the search for local LyC leakers, the data reduction has turned out to be non-trivial and few results have been published on the subject. A first investigation of the local galaxy Mrk 54 resulted only in an upper limit to the absolute LyC escape fraction $f_{esc}$$<$ 6$\%$ \citep{2001A&A...375..805D}. In \citet{2006A&A...448..513B}, the first actual detection of LyC escape was reported for Haro 11 to be $f_{esc}$=4$-$10 $\%$. However, in \citet{2007ApJ...668..891G}, the same data set was reduced with a later version of the FUSE pipeline, and here they arrived at contradicting results with $f_{esc}$ $\lesssim$ 2 $\%$. The LyC escape was studied in eight more galaxies from the FUSE archive in \citet{2009ApJS..181..272G}. None were found to have any significant excess in the LyC, and several display a negative flux in the relevant wavelength region as a result of the over-subtraction of the reduction pipeline background.

In this article, we present a detailed analysis of the same Haro 11 data set as used by Bergvall and Grimes, to determine whether Haro 11 is a LyC leaking galaxy. In addition, we analyze the final version of the FUSE reduction pipeline, CalFUSE v3.2, itself in terms of its treatment of faint targets at the shortest wavelengths. We introduce our own background fitting procedure and an alternative way of extracting low signal-to-noise ratio (hereafter S/N) FUSE spectra. 

\section{Haro 11}
\label{section:target}
Haro 11, also known as ESO 350-IG38, is a local blue compact galaxy with M$_B$ =-20.3. It seems to share many properties with star-forming galaxies at z$\sim$3 (Lyman break galaxies, LBGs), although being at the bright end of the far-UV luminosity distribution (e.g. \citealt{2008ApJ...677...37O}). Haro 11 has a complex morphology with three main star-forming knots and seems to be dynamically unrelaxed, remniscent of a merger event \citep{2001A&A...374..800O}. The knots each contain numerous super star clusters, for which \citet{2010MNRAS.tmp..949A} derived young ages, between 1 and 40 Myr and masses between 10$^3$ and 10$^7$ M$_{\sun}$. Haro 11 is also found to have a low metallicity, with an oxygen abundance 12+log(O$/$H)=7.9 \citep{2002A&A...390..891B}, and a surprisingly low \ion{H I} content. The estimated upper limit to its \ion{H}{i} mass was found to be $\cal M$(H I) $<$ 10$^8$ M$_{\sun}$  in \citet{2006A&A...448..513B}, who also noted that Haro 11 seems to contain very extended photodissociation regions (with H$_2$ molecules) of a total mass exceeding the upper H I mass limit. The same authors also found a similar amount of ionized gas in Haro 11. Thus the dominating hydrogen phases could be H$^{+}$ and H$_2$, providing more favourable conditions for escape than in normal galaxies where \ion{H I} serves as an effective shield for the LyC radiation. \citet{2006A&A...448..513B} also investigated the ionizing stellar population and found signatures of Wolf Rayet stars in the optical as well as P Cygni profiles in the far-UV, revealing the presence of O supergiants and an ongoing burst or a burst that has only very recently ended, in agreement with the ages derived in \citet{2010MNRAS.tmp..949A}.

\begin{figure}[t!]
\centering
\resizebox{\hsize}{!}{\includegraphics{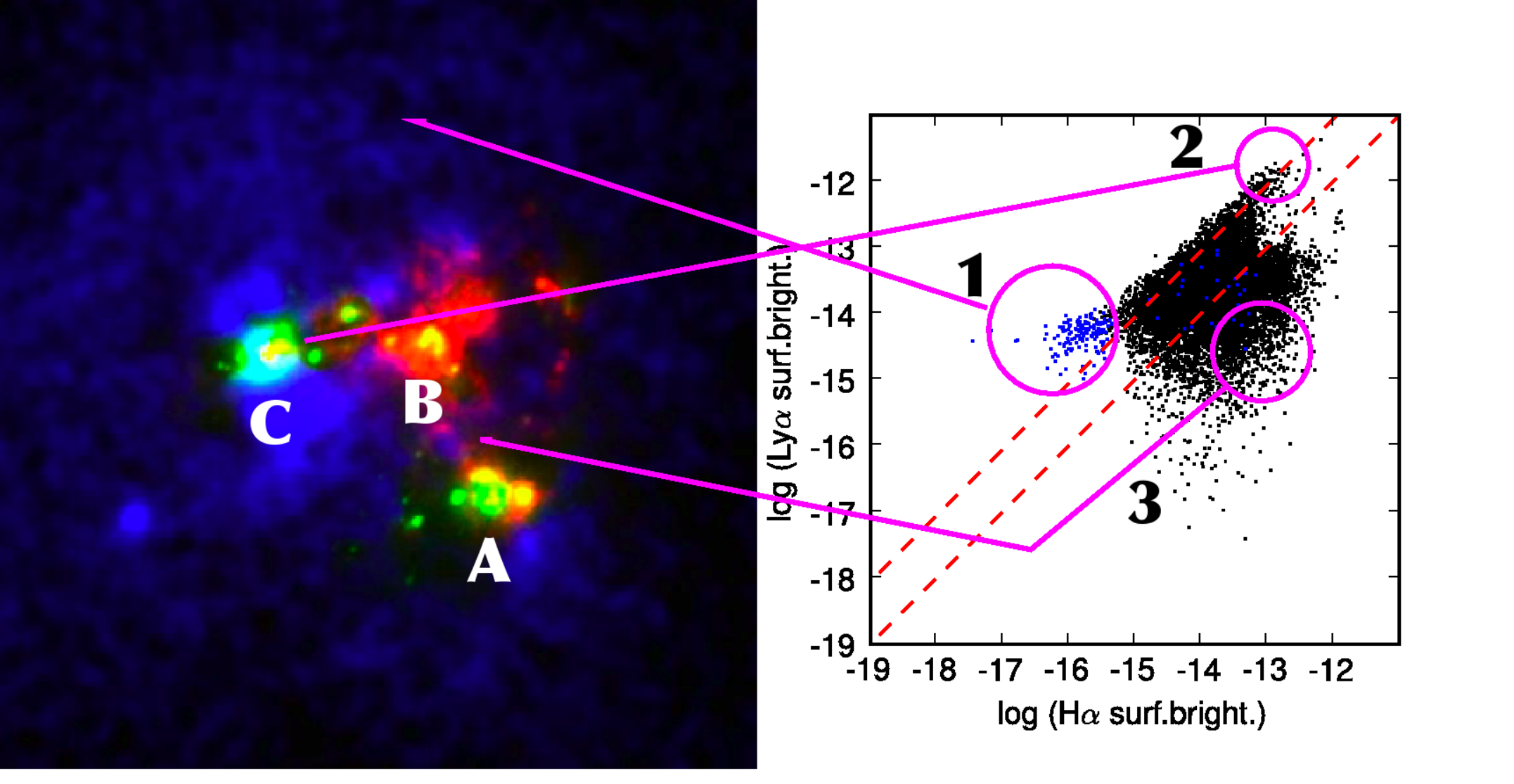}}
 \caption{\textbf{Left:} Composite HST image of Haro 11. Image scale is 15$\times$15$\arcsec$. East is left, north is up. The Ly$\alpha$ emission is shown in blue, the ionized gas via H$\alpha$ in red and the UV-continuum in green. The three main star-forming knots A, B and C are labelled. \textbf{Right:} The Ly$\alpha$ vs. the H$\alpha$ surface brightness for the binned elements in the image. The hatched lines show Ly$\alpha$=H$\alpha$ and the case B Ly$\alpha$/H$\alpha$=8.7. At low surface brightness (1) the theoretical ratio is exceeded due to resonant scattering making the Ly$\alpha$ emission off-set from the ionized regions, while at higher surface brightness (2) the top edge of the distribution agrees with the recombination value, indicating that the Ly$\alpha$ photons from these regions are scattered just a few times before escaping. Regions with heavy extinction can also be seen at high H$\alpha$ surface brightnesses (3). 
 The figure is an altered composite of Fig. 3 and 4 in \citet{2009AJ....138..923O} (see this article for further information).}
(A color version of this figure is available in the online journal.)
 \label{fig:pics}
\end{figure}
Haro 11 is also known to be a Ly$\alpha$ emitter (e.g. \citep{1998A&A...334...11K}, and the location where the Ly$\alpha$ photons escape the galaxy has been established using high resolution imaging with the ACS/SBC onboard the HST spacecraft \citep{2009AJ....138..923O,2007MNRAS.382.1465H}. These studies found that the main part of the Ly$\alpha$ photons are emitted in a diffuse halo after multiple scatterings in the interstellar medium, that is clearly offset from the ionized gas surrounding the star-forming regions (see Fig.~\ref{fig:pics} where the Ly$\alpha$ is shown in blue, the ionized gas via H$\alpha$ in red, and the UV-continuum in green). Of particular interest in the light of possible LyC leakage is one of the star-forming regions, knot C. The right panel in Fig.~\ref{fig:pics} shows the Ly$\alpha$ versus the H$\alpha$ surface brightness for binned elements. For higher surface brightnesses in knot C, it was found that the Ly$\alpha/$H$\alpha$ ratio agrees with the theoretical case B recombination value, Ly$\alpha/$H$\alpha$=8.7 \citep{2006agna.book.....O}. This indicates that the Ly$\alpha$ photons from this region are scattered just a few times before escaping. Such direct escape would be possible from a low-density, dust-poor nebula bordering on the optically thin case A,  an environment that could also allow ionizing photons to escape. If the Ly$\alpha$ photons escaped by means of an outflowing neutral medium on the other hand, as observed in \citet{1998A&A...334...11K},  such an escape mechanism would not aid the LyC photon escape unless the wind is strong enough to clear channels.

Knot C is certainly interesting in the light of possible leaking LyC radiation, but might not be the most likely option since the amount of neutral gas and dust should be largest in the central parts. \citet{2009ApJ...692.1287I} and \citet{2011arXiv1102.0286N} demonstrated that LyC leakage can be found spatially off-set from the main body of the non-ionizing UV continuum and in the central parts of the galaxies (although some of these detections statistically are likely to be lower redshift contaminants). To estimate the distance from the  center at which we would expect most of the ionizing radiation to slip through a picket fence structure in Haro 11, we adopt a simple model with two components - the ionizing stars and the neutral gas. We would ideally wish to use  a map of the UV flux, combined with a high resolution \ion{H}{i} map. Since we do not have access to these data we will instead use the B band flux distribution $\Sigma_{B,0}(r)$  \citep{2002A&A...390..891B} to model the ionizing flux, assuming a thin disk.  Since \ion{H}{i} has not been detected in Haro 11, we assume that the hydrogen gas is distributed in a homogeneous exponential disk with scalelength $h_{HI}$. In a picket-fence model, the LyC photons are absorbed by spherical clouds with radius R and  cross-section $\sigma = \pi R^2$. Thus the filling factor is $\alpha = n\cdot \frac{4\pi R^3}{3}$, where n is the number of clouds per volume unit. The observed LyC flux is then

\begin{eqnarray}
f_{LyC,obs} = 2\pi \int_0^{\infty} \Sigma_{LyC,0}(r) \cdot r e^{-\tau(r)} dr ,
\end{eqnarray}

\noindent where $\Sigma_{LyC,0}$(r) is the surface brightness in the LyC (here assumed to be $\propto \Sigma_{B,0}$), and $r$ is the distance from the centre. The optical depth, $\tau$(r) is

\begin{eqnarray}
 \tau(r) & = & \sigma \int_0^\infty n(z,r) dz \propto  \sigma e^{-r/h_{HI,r}} \int_0^\infty e^{-z/h_{HI,z}} dz   \nonumber \\ 
 && =   \sigma e^{-r/h_{HI,r}} h_{HI,z} \propto \sigma e^{-r/h_{HI,r}} ,
\end{eqnarray}

\noindent where z is the vertical height, and h$_{HI,r}$ and h$_{HI,z}$ are the radial and vertical scalelengths of the gaseous disk.

We do not know the scalelength of the \ion{H}{i} disk but a reasonable guess is that it is roughly twice the optical scale length \citep{2001AJ....122..121V,1994A&AS..107..129B,1997A&A...324..877B}, i.e. about 6 kpc \citep{2002A&A...390..891B}. We regulate $\sigma$ until we obtain a leakage corresponding to 3\%. We then find that the galaxy is leaking almost equal amounts at radii  in the range 1-14 kpc and that $\sim$50\% of the radiation is leaking through outside $\sim$7 kpc (or 17$\arcsec$) from the center. This simple model indicates that Haro 11 is an extended source in the LyC as seen by FUSE, and that we cannot rely only on the height of the spectrum longwards of the LyC to make a qualified guess  as to whether the source is extended.

\begin{figure}[t!]
\centering
\resizebox{\hsize}{!}{\includegraphics{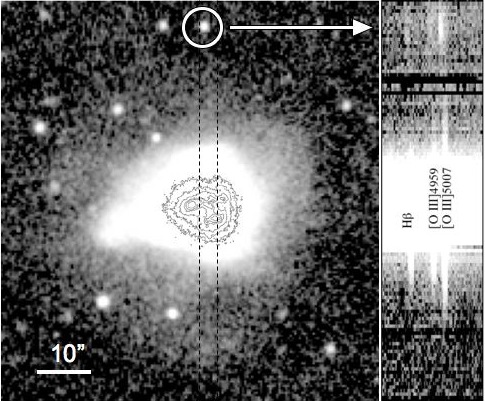}}
 \caption{Haro 11 in a deep Gunn r image. East is left and north is up. The contour plots show the far-UV continuum at Ly$\alpha$, and  help indicate the size of Fig.~\ref{fig:pics}. The slit spectrum to the right was obtained at the position marked by the hatched lines in the image. A star-forming region is found  located 35$\arcsec$ from the center north of Haro 11, illustrating that these regions exists out to, and beyond, the 30$\arcsec$$\times$30$\arcsec$ field of view covered by the FUSE LWRS aperture.}
 \label{fig:rband}
\end{figure}

In Fig.~\ref{fig:rband}, a Gunn r-band image of Haro 11 is displayed (thus including nebular H$\alpha$ emission). A slit spectrum is shown to the right, taken at the position marked by the hatched lines in the image. The image was obtained with the Gunn r filter at the 1.5m  Danish telescope at La Silla (90 minutes, 1988) and the slit spectrum with grism B300 using EFOSC1 at ESO 3.6m, La Silla (700s, 1984). In the central parts, contour plots of the far-UV continuum at Ly$\alpha$ are shown, which have a size corresponding to Fig.~\ref{fig:pics} (credit: \citet{2009AJ....138..911H,2009AJ....138..923O}). The optical size of Haro 11 is obviously much larger than the HST/ACS image shows. The slit spectrum reveals a highly blueshifted star-forming region at a distance of 9 kpc  (35$\arcsec$) . The stellar continuum in the optical is not visible, and the region is identified via its nebular emission. Although this object falls outside the FUSE aperture, it shows that star-forming regions do exist out to large distances where it might be easier for LyC photons to escape.

The redshift of Haro 11 is $z$=0.021, which places the Lyman limit (Ly-limit) at 930.6 {\AA}.

\section{Observations, reductions, and data analysis}
\label{section:obs}

The Far Ultraviolet Spectroscopic Explorer  (FUSE) is a space telescope that operated between the years from 1999 to 2007. It was equipped with a far-UV spectrometer that covers the wavelength range 905-1187 {\AA} with a resolving power of $\sim$ 20000. The four co-aligned telescopes each has a spectrograph built according to a Rowland circle design, where the spectra are imaged on two microchannel plate detectors. To accomplish good throughput over the full wavelength range, the FUSE instrument was divided into complementary channels based on  silicon carbide (SiC) optics for wavelengths shorter than $\sim$1000 {\AA} and lithium fluoride (LiF) over-coated aluminium for wavelengths in the range $\sim$1000-1200 {\AA}. Each detector is divided into two segments  that each image both a SiC and a LiF spectrum, so there are in total eight spectra. Two of these spectra cover the wavelength region where we wish to measure any possible LyC signal, SiC 1B (905-992 $\AA$) and SiC 2A (916-1005 $\AA$). For details about the on-orbit performance of the FUSE instrument, we refer to \citet{2000ApJ...538L...7S}. 

Detailed information about the satellite and its instrument can be found in the \textit{FUSE Archival Instrument
Handbook}\footnote{http://archive.stsci.edu/fuse/ih.html} and \textit{The FUSE Archival Data Handbook}\footnote{http://archive.stsci.edu/fuse/dh.html}. These two handbooks will be referred to as \textit{IH} and \textit{DH} in the text.

The observations of Haro 11 were performed on October 12, 2001, using the FUSE low resolution mode, LWRS, which has an aperture of 30$\arcsec$$\times$30$\arcsec$. The total exposure time was 16.5 ks, of which 12.2 ks was taken during orbital night. The OBSID of the data set is B1090101.

\subsection{Reductions and data analysis}
\label{section:red}

\subsubsection{The reduction pipeline, CalFUSE}
\label{section:calfuse}
We now present a short introduction to the data reduction pipeline of FUSE, CalFUSE. CalFUSE has been under constant development since the start of the mission in 1999. Although the two previous articles on Haro 11 used earlier versions of the pipeline, here we only used the final version, CalFUSE v3.2, which is described in detail in \citet{2007PASP..119..527D}. 

Running CalFUSE v3.2, the data are corrected for various motions and distortions introduced by the satellite and instrument. The arrival time and position of the photon events are preserved for time-tag observations (TTAG), and stored in two-dimensional (2D) intermediate data files (IDF). The photon list is kept until the last possible moment when the one-dimensional (1D) spectra are extracted, which makes it possible for the user to easily screen out data, e.g. for observations too close to the earth limb, during count rate bursts, during orbital day, or for photons corresponding to wavelength regions affected by airglow. All photons are flagged by the pipeline, and the user can choose to filter out the unwanted ones. 

The background model of the pipeline is of special interest in this article. The model consists of three different parts, one spatially uniform originating from intrinsic $^{40}$K decay in the microchannel plates together with cosmic rays, and two spatially varying day- and night components from scattered light. A detailed description of the internal detector background can be found in \textit{IH}, Section 4.4.3.1 and of the stray and scattered light backgrounds in \textit{DH}, Section 7.2.1. The pipeline first scales the intrinsic background and then fits fixed background files to the varying day- and night components. These template background files were produced by taking long exposures of a blank field of the sky repeatedly over the mission, and the two closest in time to the observation are selected and interpolated. The shape of the background is therefore pre-determined from those template files. This procedure works satisfactorily for bright targets, but for low S/N spectra the pipeline has been known to overestimate the background at shorter wavelengths producing an unphysical negative flux (seen in e.g. \citealt{2007ApJ...668..891G}, \citealt{2009ApJS..181..272G}). Particularly problematic are the SiC 1B and SiC 1A LWRS spectra, because they fall close to the edge of the detectors where the background has a steep slope and is difficult to model properly. A 10 $\%$ uncertainty estimate in the background was made for earlier versions of the pipeline and propagated through to the final spectrum, but in later versions this is only treated as a systematic error and as such not included in the final errors \citep{2007PASP..119..527D}. 

In several previous works where FUSE data of low S/N were treated, non-standard procedures were applied in the data reduction. While excluding orbital day data to minimize the scattered light component seems to be common in most of these investigations, they differ in other aspects. In \citet{2006A&A...455...91F}, the pulse height range was modified to reduce the background, while in \citet{2006ApJ...647..328D} they kept the standard pulse height limits with the argument of preserving photometric accuracy. In \citet{2004ApJ...605..631Z}, they extend the analysis and reduce the (night only) data with both the standard pipeline procedure and a manual extraction. In the manual method, two background regions were defined, one on each side of the spectrum, and the interpolation of the background between those made under the assumption of linear variation. For the spectra close to the detector edges, two regions were defined on one side of the spectrum and the background extrapolated from these. They found that the manual approach was more successful, as judged by the level of the residual flux in saturated absorption lines.

\subsubsection{\citet{2006A&A...448..513B} and \citet{2007ApJ...668..891G}}
There have been two previous papers published on the leakage of ionizing photons from Haro 11 based on the FUSE data. Here, a brief summary of the two is presented.

In \citet{2006A&A...448..513B}, the pipeline version CalFUSE v3.0 was used with default parameters. Haro 11 was treated as an \textit{extended source}, and the data included both the orbital day and night observations. \citet{2006A&A...448..513B} detected a visible LyC signal in both the SiC 1B and SiC 2A IDFs. The extraction of the 1D spectrum by the pipeline showed a difference between the SiC 1B and the SiC 2A flux density, and the authors corrected this by adapting the background level of SiC 1B to the SiC 2A. The LyC signal at rest wavelength 900 {$\AA$} was determined to be $f_{900}$=1.1$\pm$0.1$\times$10$^{-14}$ erg cm$^{-2}$ s$^{-1}$ \AA$^{-1}$ after correcting for Milky Way extinction. This gave an absolute escape fraction of $f_{esc}$=4$-$10$\%$, depending on the choice of parameters to describe the spectral evolutionary distribution. The authors pointed out, however,  that the uncertainty in the signal could be significantly larger than the formal error because of the complexity in determining the background (worst case 50 $\%$).

In \cite{2007ApJ...668..891G}, the same data set was reduced using CalFUSE v3.1.8. The authors here chose to treat Haro 11 as a point source, after they were certain that all the signal was included in the smaller aperture. The method to use the \textit{point source} option can be well motivated also for galaxies, since this works to improve the S/N in two ways. Not only are the data corrected for astigmatism, but the smaller extraction region also minimizes the error introduced by the background model. To reduce the contribution from airglow lines and solar scattered light, only the night portion of the data was used, reducing the exposure time from 16 ks to 12 ks. The IDF-files from the different exposures were cross-correlated with each other and coadded before the background fit and extraction of the 1D spectrum. No LyC signal could be seen by visual inspection for either 2D detector segment. The extracted 1D spectrum gave a negative flux in the LyC for the SiC 1B spectrum, but a weak positive flux in the SiC 2A \citep[see][Fig. 7]{2007ApJ...668..891G}. The authors derived an upper limit to the escape fraction of $f_{esc}\lesssim$2$\%$.

\subsubsection{Analysis of CalFUSE v3.2}
Since the two previous articles yielded such different results, and considering the known problem of CalFUSE often over-estimating the background at short wavelengths for faint targets, a detailed analysis of the Haro 11 data set reduced with CalFUSE v3.2 was undertaken. The analysis is not a comparison of the two previous articles, but instead describes the performance of the final version of CalFUSE when applied to low S/N data using different settings such as point source, extended source, normal extraction, and optimal extraction. The analysis is presented in Appendix~\ref{section:newred}\footnote{The Appendix is only available in the online journal.}. We also discuss whether Haro 11 should be treated as an extended or point source, and take a look at the special case where the signal might originate from off-center regions. 

Since the analysis is quite technical, we only present the main conclusions here and refer to Appendix~\ref{section:newred} for details.\\
\textit{i)} We cannot rule out that Haro 11 might exceed the limits defined for an object to be treated as a point source. We know that star-forming regions exists out to large distances, even outside the LWRS aperture (Fig.~\ref{fig:rband}). In addition, the spectral height of the Haro 11 continuum is found to be quite extended, roughly half that of the filled aperture airglow emission. From Fig.~\ref{fig:sic1b}, it seems as if some of the signal actually might fall outside the \textit{point source} aperture. \\
\textit{ii}) Using the \textit{point source} option, the data are corrected for astigmatism in the optical design. This is made using a model based on a true point source and is applied to the brightest source within the aperture. When applied to a galaxy where the emission might come from several regions of varying intensity with wavelength, we have no way of knowing if this is done in a correct way. For faint targets, the correction to the SiC1B spectrum could possibly even be applied to the slope of the background itself, providing the brightest pixels in parts of the spectrum. \\
\textit{iii}) Using the \textit{point source} option, the extraction of the 1D spectrum is by default made with the optimal extraction routine, meaning that central pixels are assigned a higher weight. Since we cannot rule out that any LyC emission might originate from off-center regions, such a treatment needs to be avoided and all pixels should be given equal weight.  \\
\textit{iv}) Non-simultaneous estimates of the background are insufficient for faint targets. The template background files are seen to vary in an uncorrelated way with the Haro 11 data (Fig~\ref{fig:multi}), which can cause large uncertainties in the final measured flux. The effect is most severe at the shortest wavelengths, often resulting in an overestimated background and negative fluxes.\\
\textit{v}) Even though the \textit{extended source} option was found to be technically the more reliable option, the larger extraction region applied with this method also increases the uncertainty in the background model. 

The above conclusions found from the analysis in Appendix~\ref{section:newred}, led us to develop a new background routine for reducing the FUSE data of faint targets, and to include both \textit{point source} and \textit{extended source} reduced data in our analysis.

\subsection{A new background model}
\label{section:backmodel}
CalFUSE was not originally designed for dealing with objects with very low photon counts (Dixon 2007, private communication). The main problem lies in how the background is modeled, particularly in the SiC 1A and 1B detector segments where the spectra lie on the steeply rising background at the lower edge (see Fig.~\ref{fig:sic1b} and Fig. 4.1 in \textit{DH}). In Appendix~\ref{section:newred}, we show that non-simultaneous background estimates with a predefined shape are not applicable when dealing with low signal-to-noise data. The approach here is instead to use the more standard method of directly modeling the background from the data, which also has the advantage of dealing with features from stray and scattered light described in \textit{DH}, Section 7.2.1. 

 \begin{figure}[t!]
\resizebox{\hsize}{!}{\includegraphics{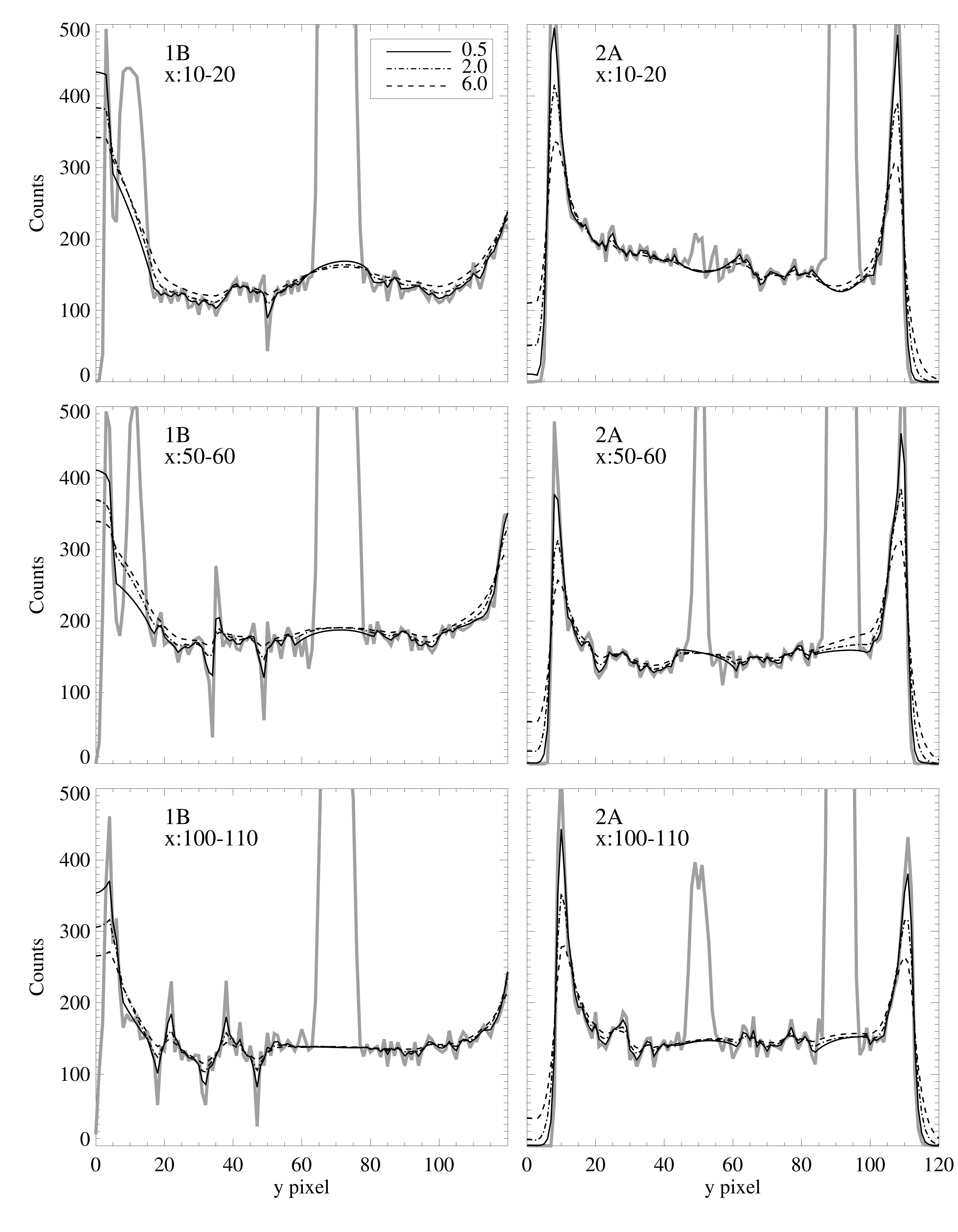}}
 \caption{The 1B (left) and 2A (right) detectors plotted in the cross-dispersion direction. The data were reduced as \textit{extended source}, and were binned with a factor 128 in x and 8 in y, to a 128$\times$128 pixel sized image. The data are summed over the pixel intervals x=10-20 (upper), x=50-60 (middle) and x=100-110 (lower). The original signal is plotted in gray, with three different background fits overplotted in black: $\Lambda_x$=$\Lambda_y$=0.5 (solid), $\Lambda_x$=$\Lambda_y$=2.0 (hatch/dot, and $\Lambda_x$=$\Lambda_y$=6.0 (hatched). The SiC spectra are centered at y pixel $\sim$10 for 1B and y pixel $\sim$ 50 for 2A, while the LiF spectra are centered at $\sim$ 70 for 1B and $\sim$ 92 for 2A. The narrow dips and peaks that are the most prominent on the 1B detector are artifacts of pipeline processing (Dixon 2007, private communication).}
 \label{fig:SiC2Aback}
\end{figure}

The new background model is constructed as follows. First we run CalFUSE v3.2 using both the \textit{extended source} and \textit{point source} options to obtain the IDF files, and combine the different exposures as recommended in \citet{2007PASP..119..527D}. Since all sub-images are co-added before the fit, we automatically deal with any time variations in the data. All the photon events are then reconstructed using the final pixel x- and y coordinates from the IDF, and the data are binned to a 128$\times$128 pixel sized image. The binning is made under the assumption of background continuity and smoothness, and allow a higher signal when creating the background model. A mask of matching size is then produced that excludes the regions where the signal lies and the edges of the detectors. Since it is difficult to achieve a good fit to the steep edges at the upper and lower parts of the detectors, and the drop-off outside these, we mask out everything outside the maximum pixel value in the edges. For segment 1B, the remaining pixels between the SiC aperture and the edge are too few in number (sometimes even zero) to be able to perform the fit. Therefore, the \textit{point source} extraction window is imported and used for defining the mask also for the \textit{extended source} spectrum (but not for the extraction of the 1D spectrum). The background is then modeled by fitting a smooth surface to the 128$\times$128 masked image using optimal filtering techniques. We achieve this by solving a regularised optimisation problem

\begin{eqnarray*}
 && \sum_{x,y} M_{x,y} \left[ D_{x,y} - F_{x,y}\right]^2+ \Lambda_x\sum_{x,y} \left[F_{x+1,y}-F_{x,y}\right]^2 \\
  & & + \Lambda_y\sum_{x,y} \left[F_{x,y+1}-F_{x,y}\right]^2 = \mathrm{min}  \\
\end{eqnarray*}

\noindent where $M_{x,y}$, $D_{x,y}$, and $F_{x,y}$ are the mask, the data, and the model value at image "pixel" $x,y$. The constants $\Lambda_x$ and $\Lambda_y$ control the smoothness of the model in the x and y directions, and are selected by the user. The model is computed by solving a sparse system of linear equations with a band-diagonal matrix. The solver is implemented as an IDL routine \textit{opt\_filter\_2d.pro} written by N. Piskunov, which is part of the \textit{Reduce} package\footnote{The \textit{Reduce} package and \textit{opt$\_$filter$\_$2d.pro} routine can be found at: \\ http://www.astro.uu.se/$\sim$piskunov/RESEARCH/REDUCE/reduce.html}. The \textit{Reduce} package is described in \citet{2002A&A...385.1095P}, but since the \textit{opt$\_$filter$\_$2d.pro} was developed later, a description of the routine was not included in this paper.

 \begin{figure}[t!]
\resizebox{\hsize}{!}{\includegraphics{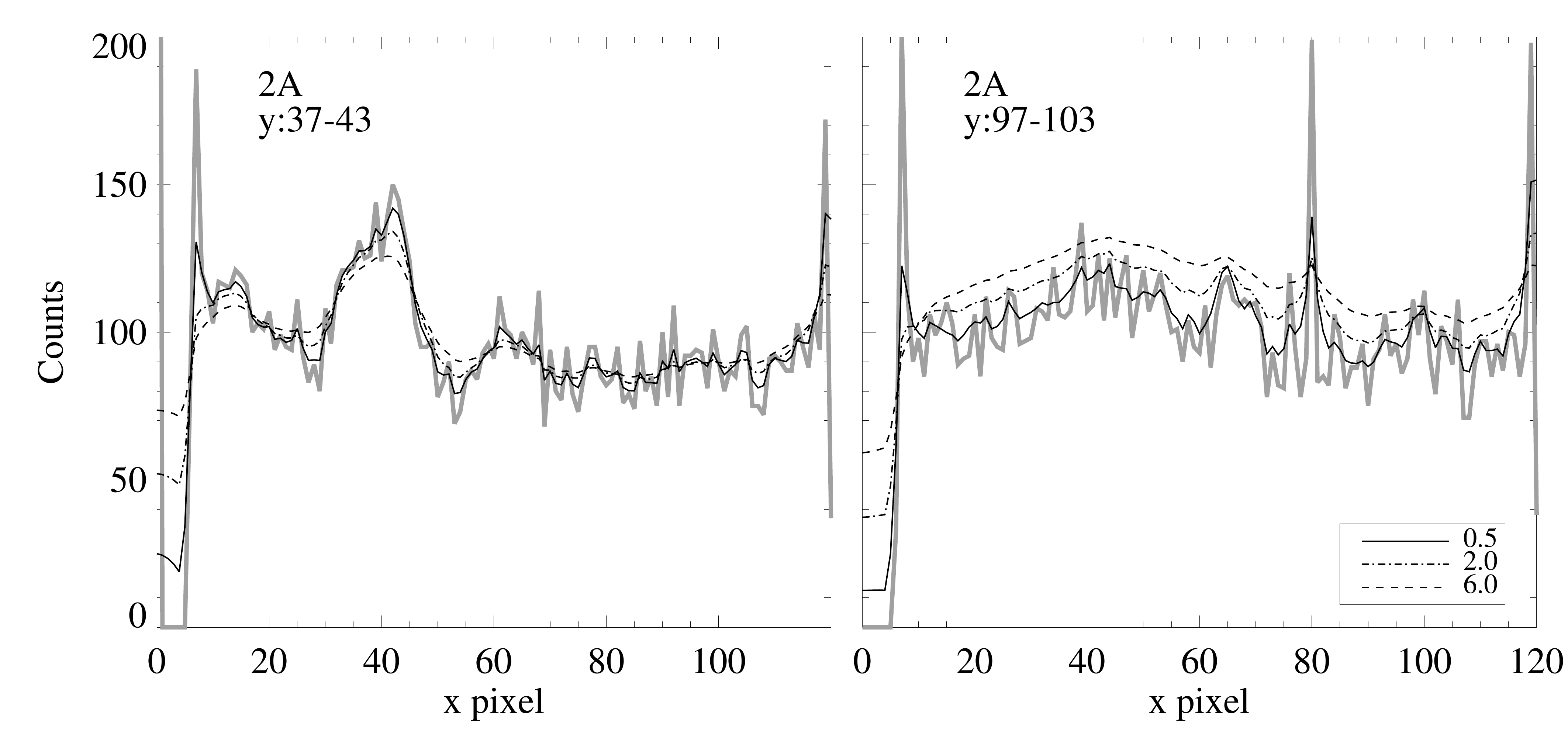}}
 \caption{The signal in the 2A detector segment plotted over the whole x range, summed over row pixels y=37-43 (left panel) and y=97-103 (right panel). The left panel samples the background and background fits just below the SiC spectrum, while the right panel samples the background and background fits just above the LiF spectrum. The original signal is plotted in gray, and three different background fits are overplotted: $\Lambda_x$=$\Lambda_y$=0.5 (solid), $\Lambda_x$=$\Lambda_y$=2.0 (hatch/dot), and $\Lambda_x$=$\Lambda_y$=6.0 (hatched).}
 \label{fig:allx}
\end{figure}

In Fig.~\ref{fig:SiC2Aback}, the background fit is shown at three different x-positions for both detector 1B (left panels) and 2A (right panels), each with three different choices of smoothing factors. The combination of $\Lambda_x$, $\Lambda_y$ that yielded the best fit to the background with avoidance of the signal areas was found to be 0.5, 0.5 for detector segment 1B, and 2.0, 2.0 for 2A. The lower smoothing factor for 1B was necessary in order to try and fit the steep slope of the background in the SiC spectrum, but even using this loose constraint the fit failed to reproduce the shape to a satisfactory level. The reason for this is simply that after the mask is applied there are not enough data points left on the slope to perform the fit, even though the more narrow \textit{point source} extraction window was used. For the 2A segment on the other hand, the SiC spectrum falls on a region of the detector where the background is much more well behaved. The fit was found to be rather insensitive to the choice of smoothing parameter in the SiC spectrum, while in the LiF spectrum closer to the upper edge it is more dependent. In Fig.~\ref{fig:allx}, this is illustrated for the whole x-range for y-pixels=37:43 (just below the 2A SiC spectrum) and y-pixel=97:103 (just above the LiF 2A spectrum). The flux was calibrated using an interpolation between the two CalFUSE effective area files closest in time to the Haro 11 observations. The error in the background fit for the SiC 2A spectrum is shown as the hatched lines in Fig.~\ref{fig:1D1}, and the error is propagated through to the final 1D spectrum and the S/N calculations.

 \begin{figure}[t!]
\resizebox{\hsize}{!}{\includegraphics{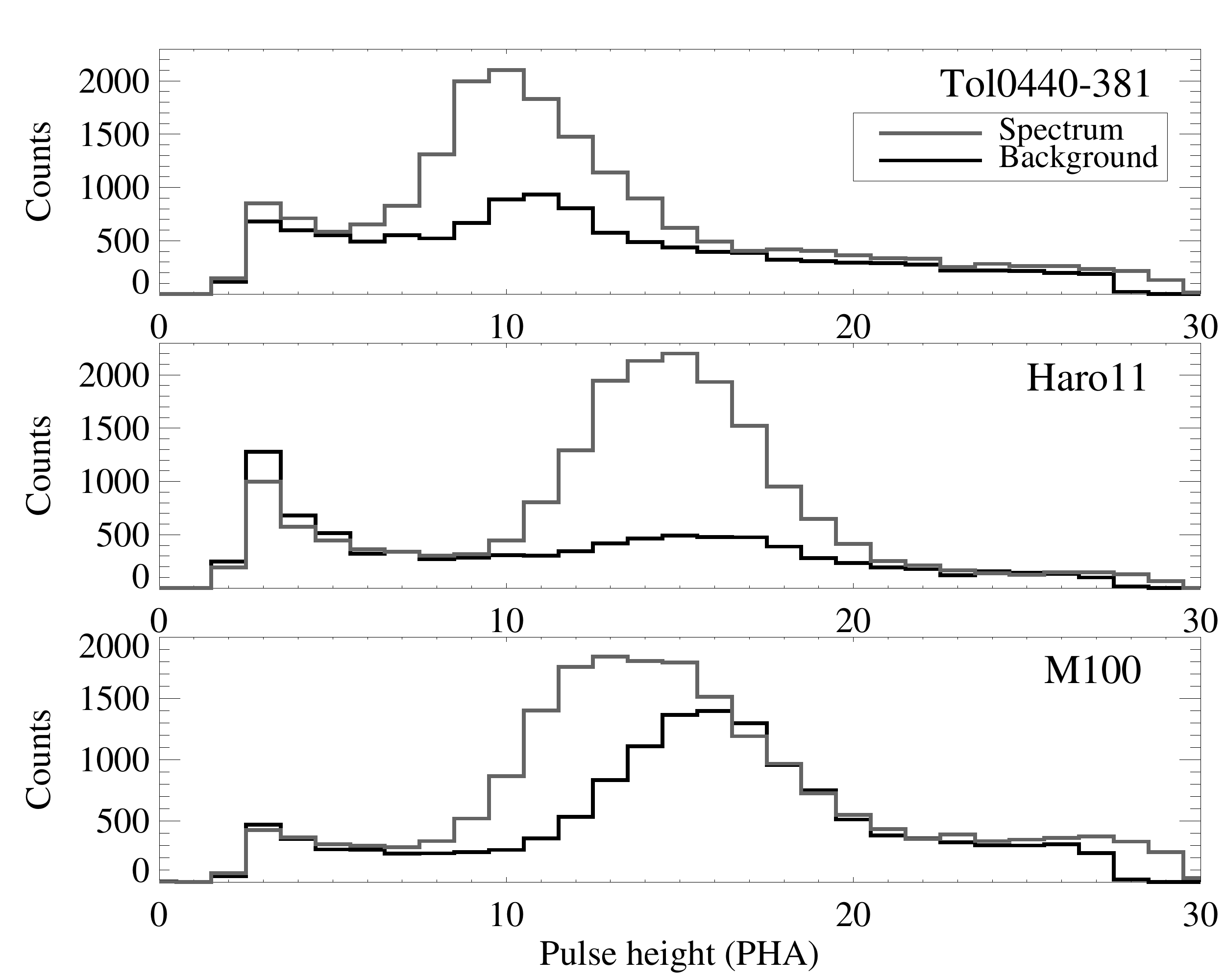}}
 \caption{The pulse height distributions for three different galaxies observed at different voltage settings throughout the mission. The pulse height distribution of the 920-980 {\AA} wavelength region in the SiC 2A spectrum is plotted in gray, with the background distributions from an equally large area under the spectrum plotted in black. Upper: Tol 0440-381, Middle: Haro 11, Lower: M 100. The drift between the two distributions show that the differential loss of sensitivity for the spectrum and background regions could have a significant effect for M 100, which was observed late in the mission after several detector voltage changes. In addition, for Haro 11 and Tol 0440-381 there is a slight shift, indicating that there might be some effect present.}
 \label{fig:pa}
\end{figure}

Since the SiC 1B fit was found not to be sufficiently good, a different approach was tried for this segment. First we reduced the binning of the IDF data, and produced a 256$\times$256 pixel image. This made it easier to locate the separation between the slope of the background edge and the signal in the SiC LWRS spectrum. Instead of masking out the point source region as before, we instead fit the background slope under the signal with a Gaussian. In a region of four pixels over the whole x range, eight points on the slope were interactively recorded to which a fourth-order polynomial fit was applied. After visual confirmation of its closeness, a given fit solution was saved to the image from which the background was modeled using the \textit{opt$\_$filter$\_$2d.pro} routine as before ($\Lambda_x$=$\Lambda_y$=2.0). To minimize the impact of the user, nine of these backgrounds were produced from which a mean image was made. The SiC 1B \textit{extended source} 2D image summed over the LyC region with the fit overplotted is shown in the lower panel in Fig.~\ref{fig:2dsumall}. The 1D spectrum with the mean background overplotted can be seen in Fig.~\ref{fig:1D1B}.  Here, also the standard error of the background fit calculated from the nine images is displayed in lighter gray. No similar attempt to fit the background of SiC 1B with \textit{point source} reduced data was tried since the separation of the background slope and the signal is very difficult to identify, and the shape of the slope of the background itself is also hard to estimate (see Fig.~\ref{fig:sic1b}).

An effect that is not taken into account in the construction of this background model is the gain sag of the detectors, which might lead to an underestimate of the flux. Owing to the  loss of sensitivity with time and exposure, the detector voltage was increased several times during the mission resulting in a shift towards higher pulse height values (\textit{DH}, Table 7.4). The illuminated parts of the microchannel plates lose more of their efficiency over time than the unilluminated parts (\textit{IH}, Section 4.4.2.1). Since we mask out the extraction regions and fit the background to the unilluminated parts, this means that the background in our model may be somewhat overestimated. The Haro 11 data were obtained early in the mission, when this effect was not expected to be significant. To investigate the effect, we plotted the pulse height distributions of three galaxies observed with three different detector voltage settings. Tol 0440-381 (2000-12-13) was observed with the initial voltage setting, Haro 11 (2001-10-12) after the second voltage change, and M 100 (2004-02-18) late in the mission when the voltage had been changed several times. In Fig.~\ref{fig:pa}, the pulse height distributions for both the 920-980 {\AA} wavelength region in the SiC 2A spectrum (gray) and the background in an equally large area below the spectrum (black) is shown. For M 100, the shift is significant between the two distributions, indicating that the differential loss of sensitivity can have a significant effect for this target. For Haro 11 and Tol 0440-381, there is a slight shift between the two distributions, and the differential sensitivity could also have some effect for these targets. If an effect is present, it will only work in the direction of more conservative LyC flux measurements.
\section{Results}
\label{section:result}

\subsection{The Lyman continuum measurements}
\begin{figure}[t!]
\resizebox{\hsize}{!}{\includegraphics{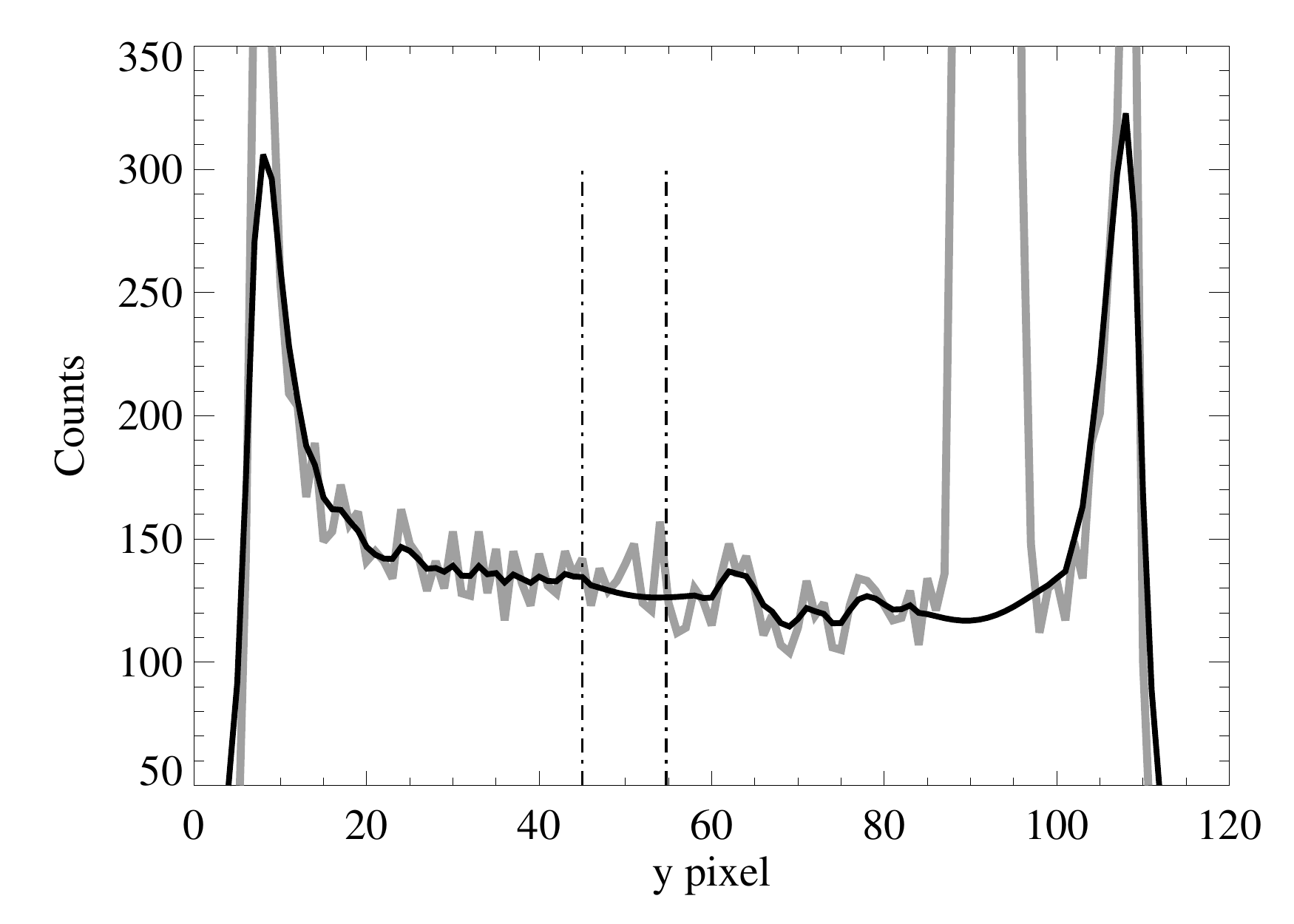}}
\resizebox{\hsize}{!}{\includegraphics{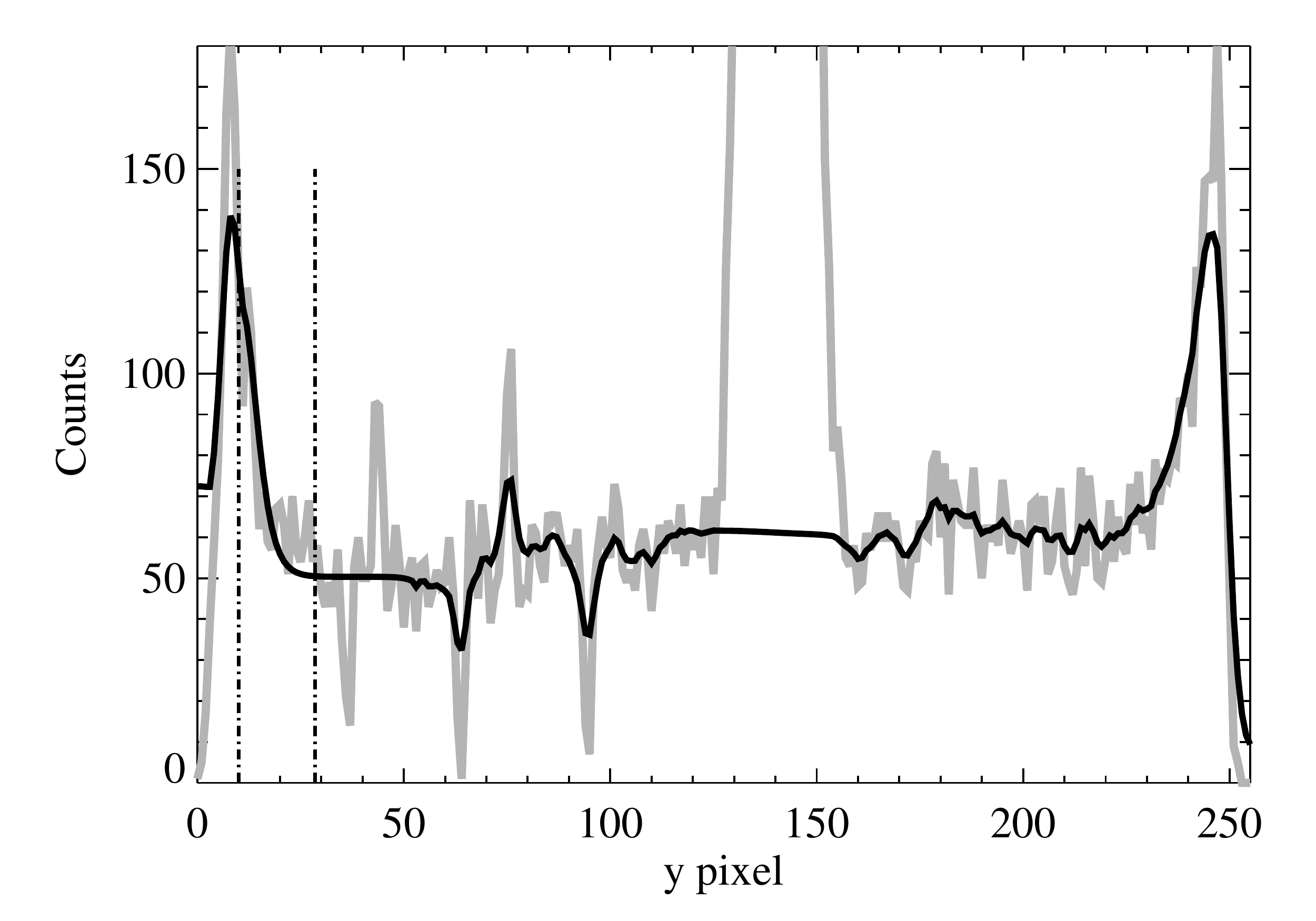}}
 \caption{The 2A (upper) and 1B (lower) segments plotted with night-only data. The signal summed over the LyC range (920-929 {\AA}) in the SiC spectrum is plotted in gray, with the new background fit overplotted in black. The vertical hatched lines indicate the extraction regions for the LWRS \textit{extended source} apertures. The SiC 2A spectrum is centered at y pixel $\sim$50, while the SiC 1B is centered at pixel $\sim$20. A weak excess can be seen in the LyC region for both spectra. }
 \label{fig:2dsumall}
\end{figure}

In Fig.~\ref{fig:2dsumall}, the night-only signal has been summed over the LyC range (920-929 \AA) in the SiC spectrum for the 2A (upper panel) and 1B (lower panel) IDFs, respectively. A weak excess can be distinguished in both spectra above the background fit plotted in black. In an effort to evaluate this signal and constrain the contributions from possible noise sources, we took a closer look at the data concerning airglow and solar scattered light, count rates, and Milky Way absorption lines.

\subsubsection{Lines affecting the Haro 11 LyC region}
Terrestrial airglow and solar scattered light emission unfortunately affects the wavelength region where we wish to measure the LyC of Haro 11. The airglow emission was studied by \citet{2001JGR...106.8119F}, using FUSE. In Fig. 7 of that article, they present spectra obtained at three different limb angles (limb, disk, and dayglow spectrum). The limb and disk exhibit distinct spectral features that are of the order of 30 times stronger than that of the dayglow spectrum. If these features are present in a spectrum it would be obvious, but neither can be identified in the Haro 11 data. Some dayglow features (airglow) are however clearly present at longer wavelengths (e.g.  \ion{O}{I} $\lambda$989{\AA} and Ly$\beta$ $\lambda$1025\AA), and in some cases in both the orbital day and night data. Although the airglow lines are weak or even invisible in the LyC wavelength region, we must consider a possible contribution from these. The airglow emission lies on top of the Milky Way Lyman absorption lines, so the emission and absorption are seen superimposed. The affected Lyman line regions have a limited extent in wavelength, so their contribution can easily be excluded when measuring the LyC of Haro 11. 

The solar scattered light is caused by solar light scattered within the SiC channel baffles (\textit{DH}, Ch 7.1). This emission can be avoided by using night data only, so the data used for the analysis in this paper should be unaffected by this scattering. By inspecting the day and night data separately, we did however locate a residual emission just below 930 {\AA}, which is likely a \ion{O}{I} airglow line. This region was excluded together with the other airglow regions. All the excluded regions are displayed as gray vertical bars in Fig.~\ref{fig:1D1}.

\subsubsection{Count rate versus limb angle}
\begin{figure}[t!]
 \resizebox{\hsize}{!}{\includegraphics{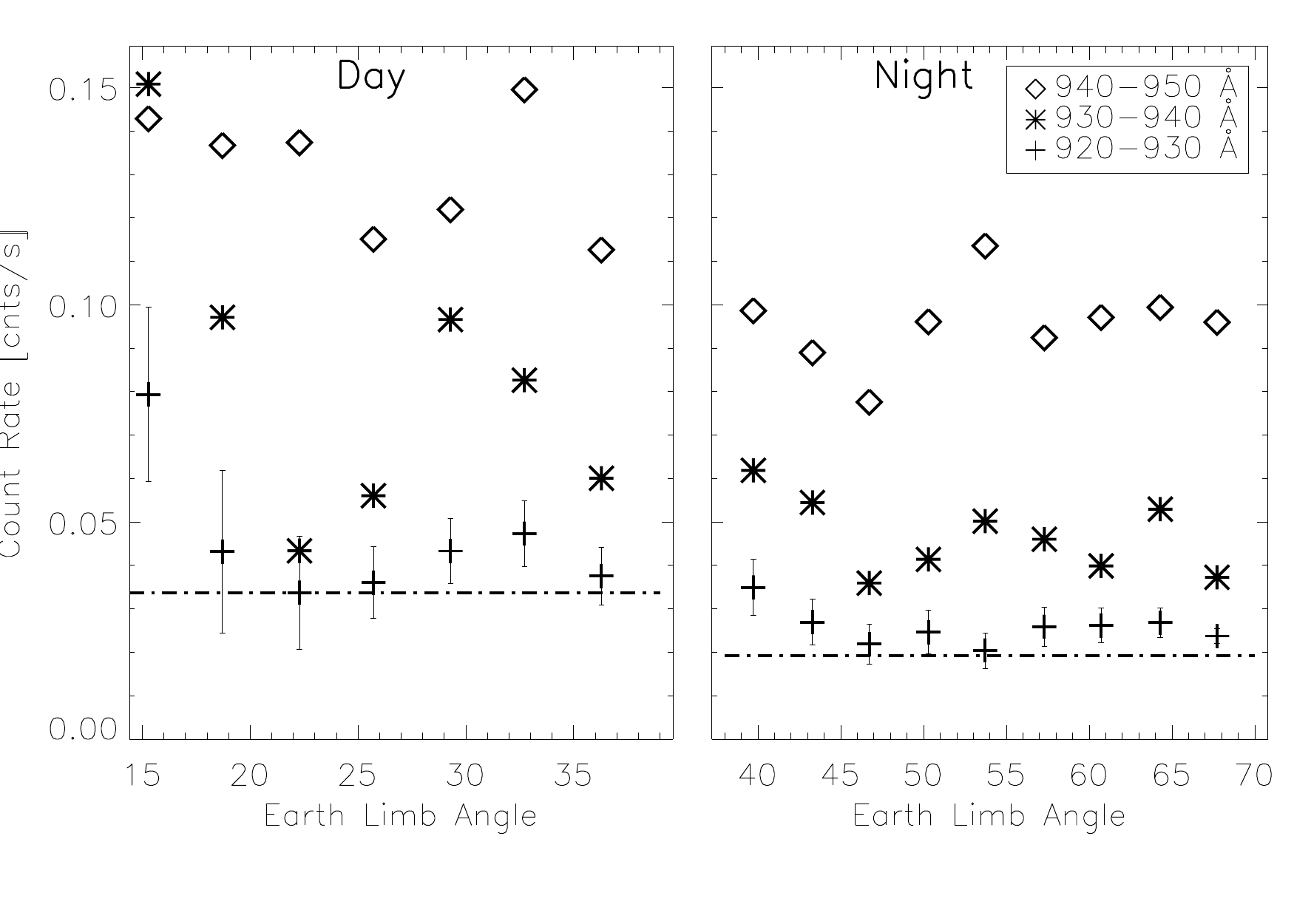}}
 \caption{The count rate for SiC 2A plotted as a function of Earth limb angle for three different wavelength windows. Airglow photons have been excluded. Orbital day data are plotted in the left panel and orbital night data in the right panel. The plots do not show any obvious trend for increasing count rate at lower limb angles neither for the day or night data, indicating that the pipeline is successful in sorting out airglow and scattered light photons. The horizontal lines show the background count rate for the 920-930 {\AA} window. A faint excess in the LyC signal above the background count rate can be seen for most limb angles.}
 \label{fig:limb}
\end{figure}

The count rate columns in the IDF files (extension 3) are produced by excluding the airglow and scattered light regions defined by the CalFUSE airglow calibration files. To check the performance of this filtering, a test was performed on the Haro 11 data. Following \citet[see][Fig. 12]{2007ApJ...668..891G}, we plotted the count rate as a function of limb angle for SiC 2A. The idea is that if the airglow and scattered light photons are successfully sorted out, the count rate will be constant with limb angle. Another interesting possibility for this plot is to see whether there is any excess in the Lyman continuum. The count rate for the 920-930 {\AA} wavelength region would in that case have to lie above the background count rate.

\citet{2007ApJ...668..891G} plotted the count rate for Haro 11 in three wavelength windows, 920-930 {\AA} (LyC region), 930-940 {\AA}, and 940-950 \AA, and found a clear trend for a decreasing count rate towards higher limb angles in all three bins. They also argue that the 920-930 {\AA} bin for high angles is close to the nominal intrinsic background count rate, implying no LyC leakage. It was unclear from their discussion why in their plot the other two wavelength windows are also close to the intrinsic value at higher angles, which is remarkable at least for the 940-950 {\AA} bin where there is a strong stellar continuum flux from Haro 11 (see  Fig.~\ref{fig:1D1}). 

In Fig.~\ref{fig:limb}, we reproduced this plot for SiC 2A. Since all the Haro 11 orbital day data were obtained for limb angles $\leq$38 degrees, and all the night data were obtained for limb angles larger than 38 degrees, we plot the count rates for these components in two different panels. As expected, there is a clear distinction between the data for the two components, there being a higher count rate overall during orbital day. From the plot, we can draw the conclusion that, if we exclude the noisy data point at the lowest limb angle, there is no clear trend for higher counts at lower limb angles for either the day or night data. This suggests that the airglow regions indeed are successfully filtered out by the calibration files of the pipeline, if that option is set by the user. The difference between day and night data can likely be attributed to a general increase in the background due to scattered light.

We also wish to explore the possibility of drawing conclusions about possible LyC leakage from this plot. The Haro 11 observations were obtained under different conditions from when the nominal intrinsic background was measured. The latter was derived at the end of the mission in 2007 with the telescope doors closed, and was averaged over the whole detector (\textit{IH}, Section 4.4.3.1). This means that the two background measurements should differ in several ways, such as stray light (which was not present when the doors were closed), the higher voltage on the detectors at the end of the mission, and the solar cycle being different (Haro 11 observations were taken well before solar minimum, while the nominal background was measured during a solar minimum). Instead of plotting the nominal background, we therefore extracted the background derived in the modeling of the Haro 11 data (Section ~\ref{section:backmodel}). The background count rate for the 920-930 {\AA} window is plotted as a horizontal hatched line in Fig.~\ref{fig:limb}. It is not a function of limb angle, but an averaged value in the 920-930 {\AA} window scaled by the exposure time and size of the region excluded due to airglow.  A weak excess in the LyC above the background can be seen for most limb angles in the night panel. The day data is only 25 $\%$ of the total observation, and the larger errors make it difficult to draw any conclusion from this panel.

We also note that the 940-950 {\AA} bin as expected has values high above the background, which in this wavelength window is 0.070 (day) and 0.047 (night) cnts s$^{-1}$.

\subsubsection{The 1D spectrum}

\begin{figure*}[ht!]
 \centering
\includegraphics[width=18.5cm]{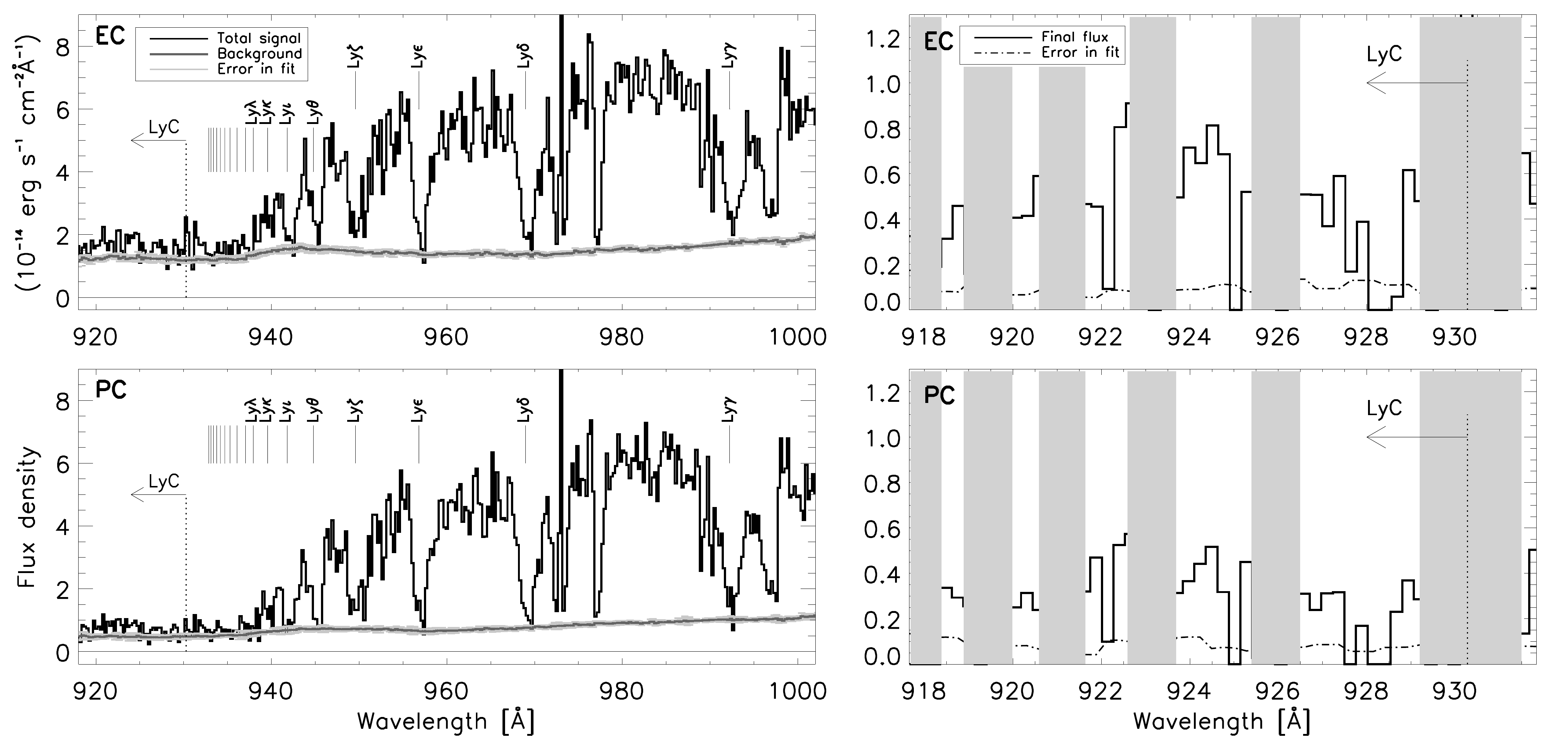}
\caption{The Haro 11 SiC 2A spectrum. The upper panels show the data reduced as extended source (EC) and the lower panels as point source (PC). The binning is 0.23 \AA. \textbf{Left:} The total signal is plotted in black, the new background in dark gray and the error in the background fit in light gray. We also indicate some intrinsic Haro 11 Lyman lines, of which only Ly$\epsilon$ at 957 {\AA} falls in a region uneffected by geocoronal or solar scattered light emission. \textbf{Right:} A closer view of the LyC region. Here, the background has been subtracted from the total signal, and the plot hence show the final flux density of Haro 11. The airglow regions are marked with gray vertical bars.}
\label{fig:1D1}
\end{figure*}

%

\begin{figure*}[t!]
 \centering
\includegraphics[width=18.5cm]{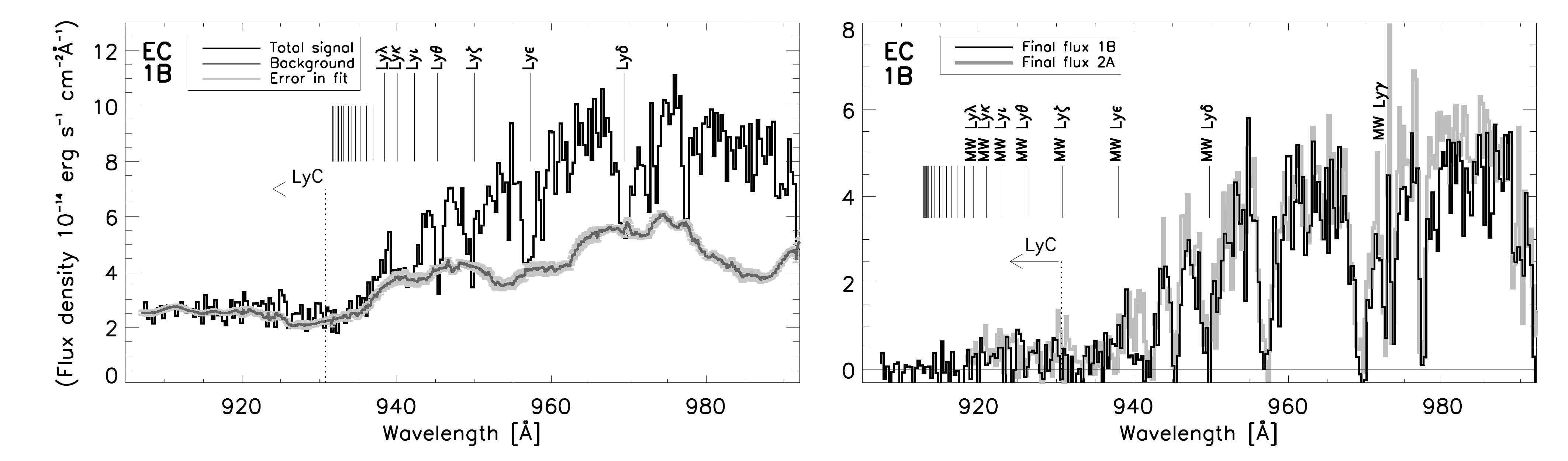}
\caption{The Haro 11 SiC 1B spectrum for the \textit{extended source} option. The binning is 0.26 \AA. \textbf{Left:} The total signal is plotted in black, the new background in dark gray and the standard error in the background in light gray. we also indicate the intrinsic Haro 11 Lyman lines down to the Lyman limit. \textbf{Right:} The final flux in SiC 1B overplotted with the final flux in SiC 2A (gray). We label the Milky Way Lyman absorption lines, for which some geocoronal emission also can be seen at the center.}
 \label{fig:1D1B}
\end{figure*}

\textbf{\textit{The SiC 2A spectrum}}\\
The 1D SiC 2A spectrum of Haro 11 using the new background model can be seen in Fig.~\ref{fig:1D1}. In the two upper panels, the data were reduced as an extended source (EC), while in the two lower panels the data were reduced as a point source (PC). The left spectra show the total signal (black), the background (dark gray), and the error in the background fit (light gray). The right spectra show a close-up of the Lyman continuum region, with the background subtracted from the total signal (i.e. the final flux density of Haro 11).  Since day data are affected by airglow and solar scattered light emission from \ion{O}{i} and \ion{H}{i} (\textit{DH}, Ch 7.1), only the night portion of the data was used. The night side \ion{H}{i} airglow regions still need to be masked out because of the presence of interstellar absorption possibly blended with geocoronal airglow emission, and are marked as gray vertical bars.

The LyC flux density ($f_{900}$) was measured on the airglow free regions below the Haro 11 Ly-limit at 930.3 {\AA}, which gave a total wavelength interval of 7 {\AA}. For the EC option, the flux density was found to be $f_{900,EC}$=4.2 $\times$ 10$^{-15}$ erg s$^{-1}$ cm$^{-2}$ {\AA}$^{-1}$ with S/N=3.8. For the PC option, the flux density was found to be $f_{900,PC}$=2.8 $\times$ 10$^{-15}$ erg s$^{-1}$ cm$^{-2}$ {\AA}$^{-1}$ with S/N=2.7. 

The Milky Way extinction at 920 {\AA} (corresponding to Haro 11 rest frame 900 {\AA}) along the line of sight is $A_{920,MW}$=0.117 \citep{2006A&A...448..513B}, and the corrected flux densities are $f_{900,EC}$=4.6 $\times$ 10$^{-15}$ and $f_{900,PC}$=3.1 $\times$ 10$^{-15}$ erg s$^{-1}$ cm$^{-2}$ {\AA}$^{-1}$.

The errors were calculated as follows. To evaluate the error in the background fit, regions above and below the SiC 2A spectrum on the 2D IDF were defined with total area equaling that of the aperture. For each resolution element the standard errors are calculated for these two regions, with respect to the original data and the background fit. The total standard error  for each airglow free window is then calculated from both the error in the background fit and the photon statistics (Gaussian) in the 1D spectrum. Finally, the weighted mean of the airglow free windows is calculated for both the signal and the error in the LyC region, where the weight is calculated from the size of the window and the effective area. In the S/N calculations the unidentified but correlated absorption features between EC and PC seen in the right panels of Fig.~\ref{fig:1D1} were treated as noise. 
\\
\textbf{\textit{The SiC 1B spectrum}}\\
In Fig.~\ref{fig:1D1B}, the 1D spectrum of SiC 1B for the \textit{extended source} option is shown. In the left panel, the total flux is plotted in black, the background in gray, and the standard error in the background in lighter gray. There is a significant difference between the background for SiC 1B and SiC 2A. While the 2A background varies little and smoothly over the whole wavelength interval, the 1B the background varies dramatically. In the right panel, the final flux (total flux minus background) is plotted in black for SiC 1B and, for comparison, the final flux in SiC 2A is overplotted in gray. The spectra seems to agree fairly well, but the flux is slightly stronger over all in 2A. It is especially interesting to see is that below $\sim$ 920 {\AA} the flux stabilizes around zero, which might be expected since the Milky Way Lyman forest  is thick at these wavelengths. An excess in the LyC is also noted, with $f_{900,1B}$=3.2 $\times$ 10$^{-15}$ erg s$^{-1}$ cm$^{-2}$ {\AA}$^{-1}$  with S/N=2.6.  Owing to the complex nature of the background surrounding the SiC 1B spectrum, the method  for calculating the errors in the background fit differs from that used to calculate the errors  in the SiC 2A fit. Here the errors in the fit are calculated as the standard error  in the nine backgrounds described in Section ~\ref{section:backmodel}.\\
\noindent
\textbf{\textit{Including day data}}\\
As a comparison, we also looked at the day+night data in the SiC 2A. The corrected fluxes were found to be $f_{900,EC}$=5.1  $\times$ 10$^{-15}$ erg s$^{-1}$ cm$^{-2}$ {\AA}$^{-1}$ with S/N=3.5, and $f_{900,PC}$=3.1 $\times$ 10$^{-15}$ erg s$^{-1}$ cm$^{-2}$ {\AA}$^{-1}$ with S/N=2.7. Both values are within one  $\sigma$ of the night-only results, indicating that scattered light during  the orbital day has a minor influence on the data.\\ \\
\noindent
\textbf{\textit{The final flux density}}\\
In the calculations of the escape fraction presented in the next section, we adopted the combined flux density from SiC 1B and 2A \emph{extended source} for night data only. The compatibility between the two spectra was evaluated on the 920-965 {\AA} region, avoiding the airglow features around 930 \AA. The null hypothesis that the two spectra are statistically comparable is supported at a confidence level of more than 90 $\%$ according to the Kolmogorov-Smirnov test, and a reduced chi-square test  of the difference between the two spectra gave a result close to unity, $\chi^2_{red}$=1.07. The final LyC flux is $f_{900}$=4.0 $\times$ 10$^{-15}$ erg s$^{-1}$ cm$^{-2}$ {\AA}$^{-1}$ (S/N=4.6).

\subsubsection{Zero flux level}
\begin{figure}[t!]
\centering    
\resizebox{\hsize}{!}{\includegraphics{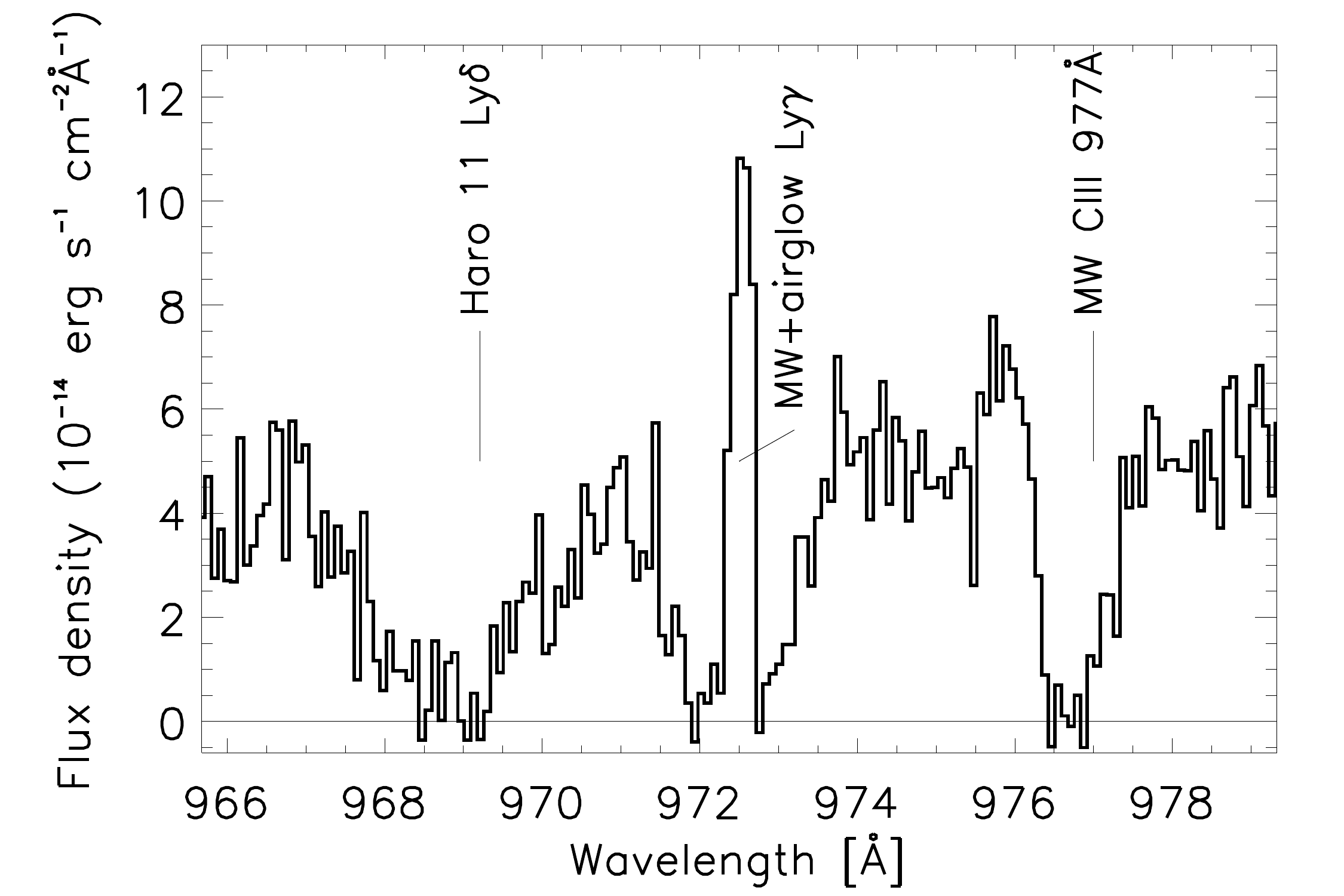}}
\caption{Part of the Haro 11 SiC 2A \textit{extended source} spectrum plotted with binning 0.08 {\AA} and night data only. The Milky Way \ion{C}{iii} $\lambda$977 {\AA} line is seen to be saturated, indicating that the zero flux level is  accurately calibrated using the background model presented in this paper.  In addition, the Haro 11 Ly$\delta$ line redshifted to 969.2 {\AA} seems to be saturated at  the line center. The Milky Way absorption Ly-lines are however affected with airglow emission, and  cannot be used as a calibration tool for Haro 11 (here Ly$\gamma$ at $\lambda$972.5{\AA}).}
\label{fig:MWlines}
\end{figure}

To validiate our results,  we need to check the zero flux level. The narrow Milky Way absorption line \ion{C}{iii} $\lambda$977{\AA} should be saturated at line center, if the line is resolved and free of solar scattered light emission (which should be absent when  night-only data are used). In Fig.\ref{fig:MWlines}, where the SiC 2A  night-only spectrum has been zoomed in on the line, it does seem as if the flux zero level is  accurately calibrated using our background model.  The intrinsic Haro 11 Ly$\delta$ redshifted to 969.2 {\AA}, also appears to be saturated at line center.

Another  possible way to calibrate the zero level was presented in \citet{2004ApJ...605..631Z}, where the faint QSO HE 2347-4342 was analyzed. Using  night-data only, saturated intergalactic \ion{He}{ii} lines were fitted with a smoothly varying polynomial to determine the zero level as a function of wavelength. For the Haro 11 data set however,  this approach was not an option since the Ly-lines are  also affected by airglow when  night-only data are analyzed. In Fig.\ref{fig:MWlines}, both the airglow emission and the broader interstellar absorption can be seen for Ly$\gamma$ at $\lambda$972.5{\AA}. The airglow emission is observed down to $\sim$930 {\AA}, and below that no clear lines can be observed indicating that the airglow emission and interstellar absorption are well blended and sometimes even cancel each other. 

\subsection{The LyC escape fraction}
\label{section:rest_escape}
Following \citet{2001A&A...375..805D}, the absolute escape fraction $f_{esc}$ is defined as the  ratio of the observed flux density of 900 {\AA} photons escaping the galaxy without being absorbed, $f_{900,obs}$, to the 900 {\AA} flux density actually produced by the stars in the galaxy, $f_{900,int}$,  given by

\begin{equation}
	\label{equation:fesc}
	f_{esc}=\frac{f_{900,obs}}{f_{900,int}} .
\end{equation}

\noindent To estimate the absolute escape fraction of LyC photons, we need  to know the total number of ionizing photons produced in the galaxy. A detailed way of deriving the intrinsically produced 900 {\AA} flux for Haro 11 was presented in \citet{2007MNRAS.382.1465H}. With high resolution HST images, they  were able to model the $f_{900,int}$ at each binned pixel based on the normalization of the  best-fit population synthesis models. Integrating the ionizing flux over the whole galaxy, they derived a total flux density of $f_{900,int}$=12.3 $\times$ 10$^{-14}$ erg s$^{-1}$ cm$^{-2}$ {\AA}$^{-1}$. Using this value, we derive an escape fraction for Haro 11 of $f_{esc}$=3.3$\pm$0.7 $\%$.

\subsection{The \ion{C}{ii} $\lambda$1036{\AA} escape fraction}
\label{sect:cII}
\citet{2001ApJ...558...56H}  presented a method  for indirectly measuring the LyC escape from galaxies. The method is based on the expectation of  residual emission in the center of the \ion{C}{ii} $\lambda$1036{\AA} interstellar absorption line if the target is a LyC leaker. Five local FUSE galaxies were observed in this pilot study, and they were all found to be essentially black in the \ion{C}{ii} $\lambda$1036{\AA} line. \citet{2009ApJS..181..272G} and \citet{2011ApJ...730....5H} increased the sample size with a factor $\approx$3 using FUSE archival data. Neither found strong evidence of any large leakage, although three of the galaxies were found to have rather high upper limits, $f_{esc}$$<$10-18$\%$. 

Haro 11 was also included in the sample. The ratio of the line core  flux of \ion{C}{ii} $\lambda$1036{\AA}  to that of the continuum was found to be R$_{CII}$=8 $\%$.  Applying CalFUSE  to detector segment 1A with the background model presented in this paper, we measured a similar ratio  of R$_{CII}$=8.5 $\%$. After correction for the far-UV extinction, they arrived at an upper limit of $f_{esc,LyC}$ $<$ 0.6$\%$. At first, this may seem  to contradict our results of $f_{esc}$, but we claim that this is not necessarily the case. 

Since it has been shown that Haro 11 is optically thick at 900 {\AA} if the absorbing gas is diffusely distributed, any emitted LyC photons must escape following the picket fence model. Because of this, there are two effects  that must be considered. Firstly, there are two components in the spectrum, one that is attenuated by dust and one that is not. Differential extinction between the unattenuated Lyman continuum at 900 {\AA} and the attenuated spectrum at the wavelength used by  \citet{2009ApJS..181..272G} to calibrate the UV flux, i.e. $\lambda$=1150 {\AA},  plays an important role.  The escape fraction can only be the same at 1036 {\AA} as at 900 {\AA} if the spectral shape of the reddened part of the spectrum is the same as  that for the part leaking through the fence, which is an unlikely scenario. Following the SMC extinction curve published by \citet{2005ApJ...630..355C} and $A_{960}$=2.9$^m$ \citep{2006A&A...448..513B}, we obtain a difference in extinction between 1036 {\AA} and 900 {\AA} corresponding to a factor  of 1.3. Secondly, we are  probably detecting two different stellar populations,  one whose light is observed through the holes  and is dominated by very young stars (1-3 Myr),  the other whose light originates from the surrounding regions which contains a more mixed population. The far-UV slopes of the two populations are therefore likely to vary, and the fluxes deviate significantly when extrapolating to LyC wavelengths. Since most of  the light of Haro 11  detected in the far-UV is enclosed within the LWRS aperture, the mean age of the stars in the UV spectrum will be significantly higher than 1-3 Myr. \citet{2006A&A...448..513B} derived a  best-fit age of 14 Myr using the models of synthetic spectra for star-forming galaxies by \citet{2003ApJS..144...21R}. This must be considered as a lower limit taken into account that the leaking population has not been subtracted prior to the fit, and that the models only include spectra for stars with ages $\leq$15 Myr. In their study of young stellar clusters in Haro 11, \citet{2010MNRAS.tmp..949A} claim that the burst has lasted for 40 Myr.  

\begin{figure}[t!]
\centering
\resizebox{\hsize}{!}{\includegraphics{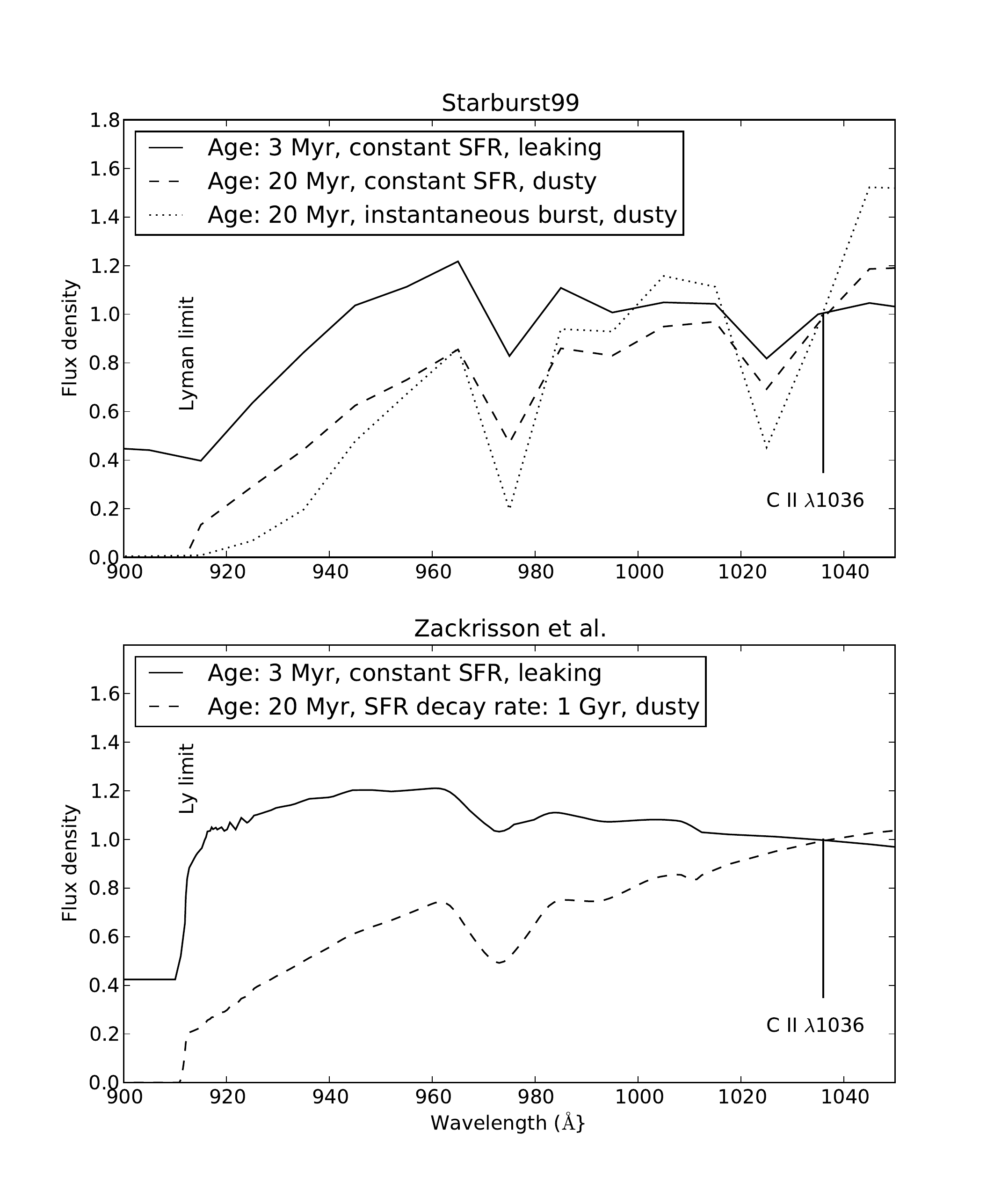}}
 \caption{Predicted spectral energy distributions in Haro 11 illustrating the picket-fence model. The stellar spectra are obtained from two models with 20 $\%$ solar metallicity - Starburst99 \citep{1999ApJS..123....3L} (upper) and our model \citep{2001A&A...375..814Z}  (lower). We assume that the leaking stellar population has a  young age, 3 Myr, has had a constant SFR and is  unaffected by extinction. The non-leaking component is assumed to be 20Myr. In the Starburst99 model, we look at two star-formation histories of this component, an instantaneous burst and a model with continuous star-formation. In the Zackrisson et al. model, we assume an exponentially declining SFR on a timescale of 1 Gyr. The old component has  been reddened by assuming the same extinction as observed in Haro 11. The Zackrisson et al. model spectra have been smoothed with a triangle smoothing to reduce the differences in spectral resolution between the stellar spectrum (low resolution) and the nebular spectrum (high resolution).}
 \label{fig:models}
\end{figure}

Two plausible models demonstrating the difference in spectral energy distribution between $\lambda$1036{\AA} and the Ly-limit with both these effects taken into account, have been summarized in Fig.~\ref{fig:models}. In the upper panel, the Starburst99  model \citep{1999ApJS..123....3L} is plotted with the leaking population (3 Myr, constant star-formation rate (SFR)) in solid line, and two non-leaking populations (hatched: 20 Myr with constant SFR, dotted: 20 Myr, instantaneous burst). The lower panel obtained with the \citet{2001A&A...375..814Z} model, essentially show the same thing but the non-leaking model is plotted for a 20 Myr population with a constantly declining SFR over  a timescale of 1 Gyr. The models are for stellar populations with 20 $\%$ solar metallicity, similar to that of Haro 11. All non-leaking models have been corrected with the same extinction as observed in Haro 11. The ratio of the leaking  to non-leaking populations is most sensitive to the extinction and the age of the leaking population. While younger stars would lead to a higher correction factor, a dominating older population would make the effect smaller. Raising the age of the non-leaking population to 50 Myr has however almost no effect at all. 

Both models in Fig.~\ref{fig:models} indicate that with the combined effects, the extinction-corrected LyC flux density derived from the \ion{C}{ii} $\lambda$1036{\AA} line should be corrected with a factor $\approx$4 from 1036 {\AA} to 900 \AA. This value was derived using a reasonable model with a leaking population  of age 3 Myr. Correcting the absolute escape fraction derived by \citet{2009ApJS..181..272G}, we arrive at $f_{esc,LyC}$ $<$ $2.4\%$. Calculating with our measured ratio R$_{CII}$=8.5 $\%$, and A$_{960}$=2.9$^m$, we arrive at the same value, 2.4 $\%$. 

\section{Summary and conclusions}
As of today, direct detection of Lyman continuum leakage has only been claimed for one galaxy in the local universe, Haro 11 \citep{2006A&A...448..513B}. In a similar analysis, using a later version of the FUSE data reduction pipeline, \citet{2007ApJ...668..891G} could not confirm  this detection. The aim of this work was to settle the question and perform a detailed analysis of Haro 11 data for the two channels covering the LyC region, SiC 1B and SiC 2A.

In previous articles where z$\sim$3 galaxies  were imaged in the LyC, it  was shown that  emitted ionizing photons can be found at  some distance from the center of the galaxies \citep{2009ApJ...692.1287I,2011arXiv1102.0286N}. While there is no  possible way of acquiring LyC images of Haro 11, we do know that this local galaxy has star-forming regions out to, and beyond, the 30$\times$30\arcsec LWRS aperture of FUSE. Even though the spectral height of Haro 11 at the wavelengths of the non-ionizing continuum was found to be  close to the size of a point source target, a simple model  indicates that the LyC emission might be seen as extended by FUSE. The analysis in this paper was therefore made with both the \textit{extended source} and \textit{point source} pipeline options. 

The detailed analysis of the reduction pipeline in Appendix~\ref{section:newred} shows that one has to be very careful using the \textit{point source} option of CalFUSE v3.2 for galaxies and potential Lyman continuum leakers where the signal might originate from off-center regions. We discuss the importance of carefully choosing the cross-dispersion centroid for the extraction regions, the risk of astigmatism corrections, and the skewed weighting of central pixels as the 1D spectra are extracted using this option. The \textit{extended source} option on the other hand, although found to technically be the more reliable option,  increases the uncertainty introduced by the background fit.

The non-simultaneous background model used by the reduction pipeline was found to be inadequate for the low  S/N spectrum of Haro 11. The shape of the template background file can be seen to vary in a way sometimes uncorrelated to the Haro 11 data, and the effect is most severe at the shortest wavelengths. A new method  for subtracting the  background was therefore developed, where the background is modeled by fitting a smooth surface directly to the detector using optimal filtering techniques. The new background model was found to be very accurate for the SiC 2A spectrum, but not for the SiC 1B spectrum positioned on the steep slope of the detector background. For this spectrum, the model had to be modified by fitting a Gaussian to the slope of the background of the target signal. No attempt to fit the SiC 1B \textit{point source} reduced data was made,  owing to the difficulty in  distinguishing between the background slope and target signal. A weak LyC excess in the 2D data above the background could be identified in the  night-only data of both SiC 1B and SiC 2A.

Plotting the count rate  versus limb angle displayed no trend, indicating that the calibration files of the pipeline successfully filters out airglow and solar scattered light photons if this option is set by the user.  In addition, by comparing the count rate in the LyC  range (920-930 {\AA}) to the corresponding background count rate, we found an excess in the LyC for most limb angles. 

 Applying CalFUSE v3.2  to the Haro 11 data set  in addition to the new background model, we extracted the 1D spectra. The LyC signal observed in the 2D data and count rate plot could  also be confirmed here. Using the \textit{extended source} SiC 1B and SiC 2A night only data, the flux density in the LyC between the airglow regions was found to be $f_{900}$=4.0 $\times$ 10$^{-15}$ (S/N=4.6) erg s$^{-1}$ cm$^{-2}$ {\AA}$^{-1}$. The errors were calculated including both photon statistics and the uncertainties in the background fit. From this, we derive an absolute escape fraction of  $f_{esc}$=3.3$\pm$0.7 $\%$.

Since it has been shown that Haro 11 is optically thick at 900 {\AA} if the absorbing gas is diffusely distributed, any emitted LyC photons must escape following the picket fence model. Previous works where the LyC flux was estimated from a residual emission in the centre of the \ion{C}{ii} $\lambda$1036{\AA} interstellar absorption line,  failed to take into account two important effects using this model.  These effects of differential extinction and the different spectral energy distributions of the leaking and non-leaking populations, should in the case of Haro 11 cause a factor  of four difference in flux between 1036{\AA}  and the Lyman limit. Thus we derive an escape fraction of $<$ 2.4 $\%$ using the \ion{C}{ii} $\lambda$1036{\AA} line, in agreement within the errors  with our measurement  obtained directly  from the Lyman continuum. The correction factor (and $f_{esc}$) will be larger if the age of the leaking population is younger than the assumed 3 Myr.  

Thus Haro 11 continues to be the only galaxy in the local universe for which escaping LyC photons  have been detected. The low escape fraction is not unexpected  by model predictions. One important parameter in these models is the mass, where most  models seem to agree on a smaller $f_{esc}$ for galaxies with  higher masses such as that of Haro 11.  In addition, the increase in metallicity with redshift and the lack of \textit{PopIII} stars greatly reduce the probability of finding large escape fractions from local galaxies.

\begin{acknowledgements} The authors would like to thank G\"oran \"Ostlin and Matthew Hayes for providing us with Fig.~\ref{fig:pics}. The authors would also like to thank the anonymous referee for the effort and expertice which helped to greatly improve the manuscript. EL would like to thank David Sahnow and W. Van Dyke Dixon from the FUSE team for their valuable help and comments on technical details regarding FUSE and CalFUSE. \\
This work was supported by the Swedish Space Board. NB also acknowledges support from the Swedish Research Council.\\
Some of the data presented in this paper were obtained from the Multimission Archive at the Space Telescope Science Institute (MAST). STScI is operated by the Association of Universities for Research in Astronomy, Inc., under NASA contract NAS5-26555. Support for MAST for non-HST data is provided by the NASA Office of Space Science via grant NAG5-7584 and by other grants and contracts.
\end{acknowledgements}

\bibliographystyle{aa} 
\bibliography{sample2} 

\appendix

\section{Using CalFUSE v3.2}
\label{section:newred}
 Performing LyC measurements  requires data acquired while working at the very limit of instrument sensitivity.  We have  therefore dedicated this appendix to a detailed analysis  of the performance of the reduction pipeline on the Haro 11 data set. Haro 11 is one of several faint FUSE targets for which the CalFUSE background subtraction has resulted in negative flux at short wavelengths, and the results found here  may be of interest  to other similar investigations. We also approach the special case  where the emission might originate from extended or off-center regions of the targets. 

\subsection{Haro 11, extended or point source?}
This is a fundamental question when the reduction approach is to be decided. From the ACS/SBC UV image in Fig.~\ref{fig:pics}, we see that the separation between the two strongest UV emitting regions is $\sim$5$\arcsec$. In the header of the Haro 11 FUSE data, the keyword APER$\_$PA=99.4, which means that the distance between the two regions are almost maximized in the cross-dispersion direction. The LWRS aperture of FUSE is 30$\arcsec$$\times$30$\arcsec$. Only if the image height is roughly half the aperture size or more, will there be a significant effect on the height of the spectral image. Therefore,  when only considering the far-UV ACS/SBC image, should Haro 11 pass as a point source. However, Fig.~\ref{fig:pics} covers a square 15$\arcsec$, and we know that star-forming regions do exist outside this from Fig.~\ref{fig:rband}.  In the Chandra image of \citet[see][Fig. 6]{2007MNRAS.382.1465H} with the far-UV contours overplotted, the X-ray emission is seen at least out to a diameter of 25$\arcsec$.

\begin{figure}[t!]
\centering    
\resizebox{\hsize}{!}{\includegraphics{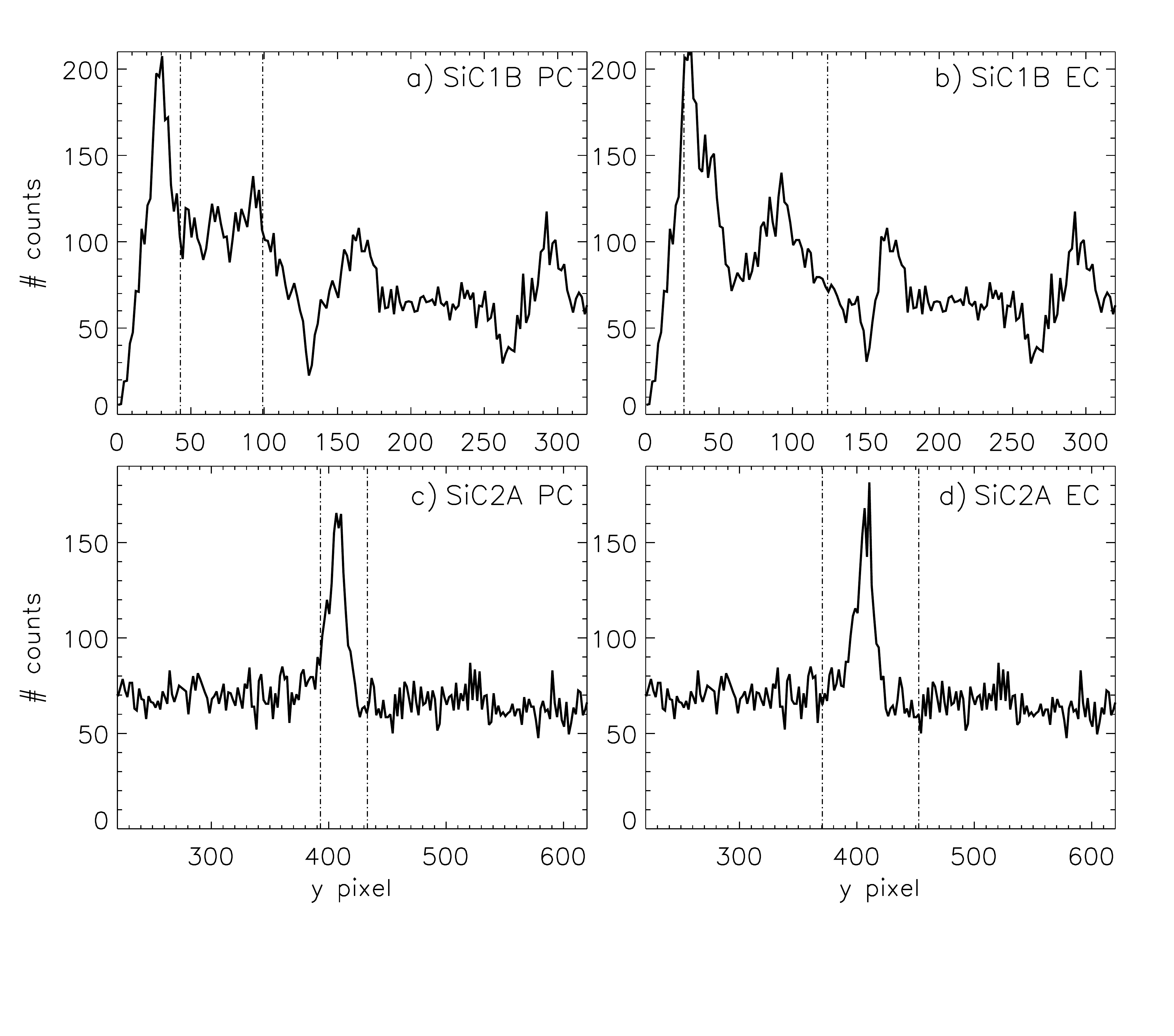}}
\caption{Part of the SiC 1B and SiC 2A IDFs seen in the cross-dispersion direction, summed over 920 - 950 {\AA}. The extraction windows for the LWRS aperture is plotted in vertical hatched lines. The cross-dispersion centroids were determined with \textit{user select} for the 1B files and \textit{target events only} for the 2A files, and the width shows the maximum width in the wavelength interval. In \textbf{a)} SiC 1B reduced using the \textit{point source} option, \textbf{b)} SiC 1B reduced using the \textit{extended source} option, in  \textbf{c)} SiC 2A reduced as \textit{point source}, and \textbf{d)} SiC 2A reduced as \textit{extended source}. 
The \textit{point source} spectra have been corrected for astigmatism. Comparing a) and b), it is clear that the effect of that is significant in the SiC 1B spectrum. It is hard to determine whether the astigmatism correction has been made on the target data or on the slope of the background. The effect on the SiC 2A spectrum is not significant. The authors are not convinced that the flux from the target lies completely within the \textit{point source} extraction windows for either 1B or 2A. The dips and peaks to the right of the extraction regions in 1B are artifacts of pipeline processing (Dixon 2007, private communication).}
\label{fig:sic1b}
\end{figure}

To check  whether Haro 11 is consistent with a point source in the spectral image of the LWRS spectrum, we performed the following test. The astigmatic image height is smallest at 916 {\AA}  in the SiC 1B spectrum and at 926 {\AA} in SiC 2A, that is,  where we wish to measure the LyC. It is also close to  reaching a minimum at Ly$\beta$ in the LiF 1A spectrum, where the continuum flux of Haro 11 is strong enough to perform the test. The airglow Ly$\beta$ fills the aperture, so if the Haro 11 continuum height is less than half of that it should pass as a point source. We find that the airglow Ly$\beta$ height is roughly 90 unbinned pixels, while the Haro 11 height just below and above Ly$\beta$  is half of that, 45 pixels. We conclude that Haro 11 is a  borderline case for being considered as a point source, and in our analysis we therefore considered both cases.

\subsection{Producing the IDFs}
Here, we reduced the Haro 11 data set with the last version of the pipeline, CalFUSE v3.2. The pipeline was run  with both the  \textit{point}- and  \textit{extended source} options in order to compare the two methods. The processing trailer files were then scanned for warnings of which none were found. The IDFs and  bad-pixel masks from the different exposures were coadded before the background fit and the extraction of the 1D spectra was made, all according to the recommended strategy for low  S/N spectra in \citet{2007PASP..119..527D}. The IDFs were then  cross-correlated before they were combined (max correction 0.026 \AA). 

\noindent \textbf{\textit{The cross-dispersion centroid}}\\
The cross-dispersion centroids of the extraction windows were calculated using the \textit{cf$\_$edit} routine with option \textit{target events only}, except for the SiC 1A and SiC 1B where the contribution from the bright detector edge tends to move the centroids. For these two, the centroids were therefore determined with the option \textit{user select} after careful inspection. This treatment turned out to be especially important for the \textit{point source} reduction. In Fig.~\ref{fig:sic1b}, the SiC 1B and SiC 2A IDFs  were summed in the cross-dispersion direction over the wavelength region 920 - 950 {\AA}. The extraction windows for the LWRS aperture are plotted as vertical hatched lines, with the width displaying the maximum width within the wavelength interval. In a) and c) where the  \textit{point source} option has been used,  part of the signal appears to fall outside the aperture. 

\begin{figure*}[t!]
\centering
\includegraphics[width=18.5cm]{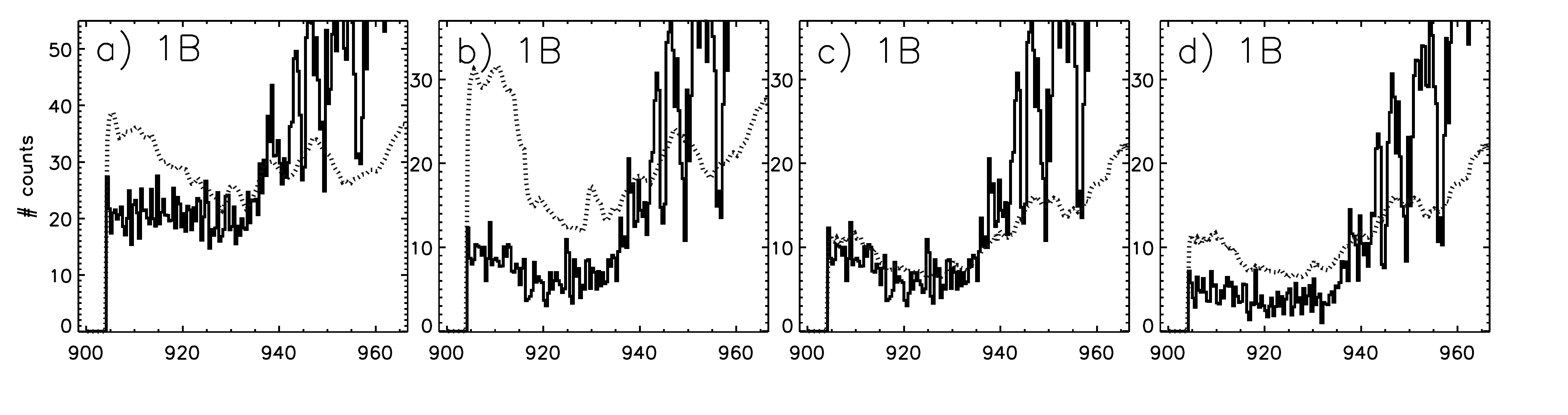}
\includegraphics[width=18.5cm]{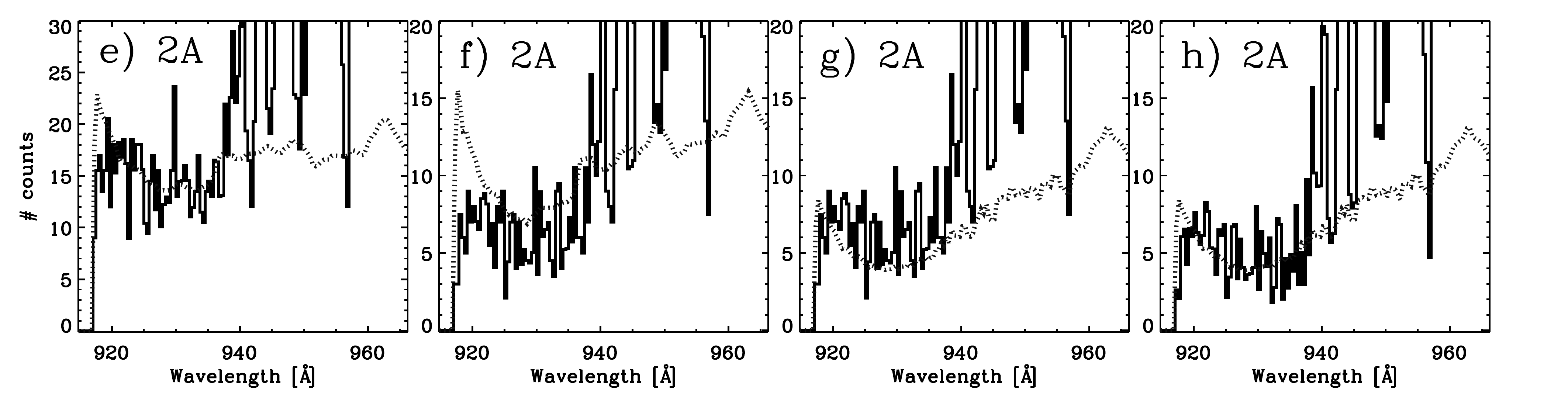}
  \caption{The plots show the extracted Haro 11 total signal (galaxy + background=\textit{WEIGHTS}) for SiC 1B (upper row) and SiC 2A (lower row), over-plotted by the fitted background (\textit{BKGD}) in dotted line. From left: \textbf{a) and e)} \textit{extended source} setting, \textbf{b) and f)} \textit{point source} setting with \textit{normal extraction}, \textbf{c) and g)} \textit{point source} setting with \textit{optimal extraction} and \textit{WEIGHTS} produced with background-subtraction routine turned off (BKGD imported from d) and h)), \textbf{d) and h)} \textit{point source} setting with \textit{optimal extraction} (corresponding to the treatment in \citet{2007ApJ...668..891G}). In theory c) should be identical to d), and g) identical to h), but they are not as discussed in the text. The spectra are rebinned by 16 pixels=0.208 {\AA}. Only night data are plotted.} 
   \label{fig:multi}
\end{figure*}

\noindent \textbf{\textit{Weights and off-center emission}}\\
In \cite{2009ApJ...692.1287I}, LyC imaging was performed  for a sample of z$\simeq$3.1 galaxies. It was found that the ionizing radiation in several cases was emitted from regions off-center from the non-ionizing emission. Looking at Haro 11, this possibility  cannot be neglected since star-forming regions are  also known to exist outside the HST/ACS far-UV image (see Sect.~\ref{section:target}), and the scenario would likely be applicable to other FUSE galaxies as well. The \textit{point source} option  that utilizes the \textit{optimal extraction} as standard routine must therefore be considered carefully when measuring possible LyC emission, not only because of the more narrow extraction window. This routine also extracts the 1D spectrum using weights, where the central pixels in the extraction window receive larger weight than the ones  closer to the edges. The \textit{extended source} option on the other hand, is not as sensitive  to the way in which the centroid is chosen,  and the 1D extraction routine is not applied using weights. 

\noindent \textbf{\textit{Astigmatism corrections}}\\
When using the \textit{point source} option, the data are corrected for astigmatism in the optical design. The model for this correction is based on a true point source, and its applicability must be questioned when considering the possibility that LyC leakage might originate from  off-center regions. The astigmatism correction is performed on the location of the brightest pixels in the aperture, disregarding the centroid of the extraction region. If the emission from different wavelength bands  originates from different spatial locations in the aperture, this will likely be done incorrectly for some of them (Sahnow 2010, private communication). In Fig.~\ref{fig:sic1b}, the danger of performing the astigmatism correction on the slope of the detector background in SiC 1B is demonstrated. It is unclear  whether the shape of what appears to be the target signal in a) is caused by a correction  to either the target signal or  the slope of the background. The SiC 2A signal on the other hand, does not seem to be affected in the same way.
\newline
\noindent  We need to consider that the LyC could be emitted from an off-center star-forming region, the uncertainty of the astigmatism correction, and the risk that not all the signal falls within the \textit{point source} extraction window,  hence we need to  consider both cases (\textit{point source} and \textit{extended source}) in our analysis.

\subsection{From 2D to 1D}
In the analysis we also wished to investigate the reliability of the 1D extraction routine, and tried different options  such as optimal- or normal extraction, and  selecting or not the background subtraction routine. There are eight 1D spectrum files produced by CalFUSE, one for each spectrum. These provide several options for the user to display the data, where for example \textit{FLUX}, \textit{COUNTS}, \textit{WEIGHTS}, and \textit{BKGD} (background) are offered. The \textit{WEIGHTS} column represents the raw counts in \textit{COUNTS} corrected for dead-time effects, but for faint targets such as Haro 11 the dead-time correction is  negligible, so the two are essentially identical. The final flux is calculated  by subtracting \textit{BKGD} from \textit{WEIGHTS}. 

The data were reduced with different parameter set-ups. First we used the \textit{extended source} setting (the default used in \citet{2006A&A...448..513B}). This can be seen in Fig~\ref{fig:multi} a), where the total SiC 1B signal (\textit{WEIGHTS}) is overplotted  with the fitted background (\textit{BKGD}, dotted line). A clear  overestimate of the background is visible at the shorter wavelengths. The same is plotted for SiC 2A in panel e). \citet{2007PASP..119..527D} argue for using  the \textit{point source} setting whenever possible in order to improve the background model for faint targets. We tried this both with \textit{normal extraction} (all pixels get equal weight) as well as  with the standard \textit{optimal extraction} routine where central pixels get more weight (the latter being the case used in \citet{2007ApJ...668..891G}). The resulting SiC 1B spectra can be seen in Fig.~\ref{fig:multi} b) - d) for 1B and f) - h) for 2A. In b) and f)  the \textit{normal extraction} produces a strong increase of the background at the shorter wavelengths. When the background is subtracted, this results not only in an unphysical negative flux, but also an artificially produced Ly-limit feature that makes it impossible to determine the zero-level of the flux below the Milky Way Ly-limit (912 \AA). In c) - d) and g) - h) the \textit{point source} setting was used with the \textit{optimal extraction} routine. The difference between them is that in c) and g) the \textit{WEIGHTS} column is taken from a file where the background routine was turned off. The overplotted backgrounds are instead taken from the files in d) and h). The \textit{WEIGHTS} columns should be identical  regardless of whether the background has been subtracted or not,  thus plot c)  is expected to be identical to d), and g) identical to h), which they clearly are not. The reason for this is not known but could possibly, according to Dixon (2007, private communication), be caused by the complex optimal extraction routine involving probability arrays used on low  S/N spectra. 

Comparing Fig~\ref{fig:multi} to Fig.~\ref{fig:1D1}, we see a large difference between the two background models in SiC 2A. In Fig.~\ref{fig:1D1} where the background has been fitted directly on the detector, the curve is smooth. In Fig~\ref{fig:multi} on the other hand, the shape of the template background file is seen to vary in a way often uncorrelated to the Haro 11 data. This demonstrates the danger of using non-simultaneous template background files, and for faint targets the effect can be severe. The features will have a large effect on the output flux for which the amplitude will vary in an unknown way over the spectrum. The effect is largest at the shortest wavelengths in all panels of Fig~\ref{fig:multi}.

\noindent \textbf{\textit{Summary Appendix A:}}\\
\noindent We can conclude that both the \textit{point source} and \textit{extended source} options must be considered for the Haro 11 data since there are benefits and risks with both. While the \textit{point source} option might be unable to include all of the signal, might skew the weighting if the emission is off-center,  and in the case of SiC 1B perform a questionable astigmatism correction,  the  \textit{extended source} option, which is otherwise more stable, has been shown to include more of the background region and therefore increase the uncertainty in the background model. We have also shown that the non-simultaneous background templates used by CalFUSE v3.2  display features  that are uncorrelated to the Haro 11 data, and that  these features can cause large errors in the final flux at random places over the spectrum (but especially at the shortest wavelengths). This analysis  has therefore led us to develop our own background model from a direct fit to the detector response (Section~\ref{section:backmodel}), which is the standard procedure when reducing astronomical data. 

\end{document}